\documentclass{article}

\usepackage{arxiv}
\usepackage[dvipsnames]{xcolor}
\usepackage[utf8]{inputenc} \usepackage[T1]{fontenc}    

\usepackage[osf]{newtxtext}         \usepackage{newtxmath}              \usepackage[cal=euler]{mathalfa}    \usepackage[scaled=0.85]{beramono}  

\usepackage{hyperref}       \usepackage{url}            \usepackage{booktabs}       \usepackage{lscape}
\usepackage[raggedrightboxes]{ragged2e}
\usepackage{multirow}
\usepackage{threeparttable} \usepackage{amsmath}
             \usepackage{amssymb}
\usepackage{mathtools}
\usepackage{nicefrac}       \usepackage[protrusion=true, expansion=true, tracking=smallcaps]{microtype}
\SetTracking{encoding=*, shape=sc}{40}  \usepackage{graphicx}
\usepackage{orcidlink}
\providecommand{\orcidlinki}[2]{#1~\orcidlink{#2}}
\providecommand{\orcidlinkf}[1]{\href{https://orcid.org/#1}{\orcidlink{#1}~\nolinkurl{orcid.org/#1}}}
\graphicspath{{graphics}}
\usepackage{csquotes,xpatch}
\usepackage[backend=biber,                          style=apa,                             uniquename=false,                       sorting=ynt,                            ]{biblatex}
\DeclareFieldFormat{doi}{\mkbibacro{DOI}\addcolon\space
  \ifhyperref
    {\href{https://doi.org/#1}{\nolinkurl{#1}}}
    {\nolinkurl{#1}}}
\usepackage[automake, nomain, acronym, xindy, shortcuts=ac]{glossaries-extra}
\glsdisablehyper
\makeglossaries
\setabbreviationstyle[acronym]{long-short} \glssetcategoryattribute{acronym}{nohyperfirst}{true}
\newacronym{aru}{ARU}{automatic recording unit}
\newacronym{ssl}{SSL}{self-supervised learning}
\newacronym{bilstm}{BiLSTM}{bidirectional long short-term memory network}
\newacronym{ast}{AST}{audio spectrogram transformer}
\newacronym{vit}{ViT}{vision transformer}
\newacronym{auc}{AUC}{area under the curve}
\newacronym{fpr}{FPR}{false positive rate}
\newacronym{tpr}{TPR}{true positive rate}
\newacronym{auc roc}{AUC ROC}{area under the receiver operating characteristic}
\newacronym{roc}{ROC}{receiver operating characteristic}
\newacronym{stft}{STFT}{short time Fourier transform}
\newacronym{fft}{FFT}{fast Fourier transform}
\newacronym{nlp}{NLP}{natural language processing}
\newacronym{cnn}{CNN}{convolutional neural network}
\newacronym{ldc}{LDC}{Linguistic Data Consortium}
\newacronym{mfcc}{MFCC}{mel frequency cepstral coefficient}
\newacronym{knn}{k-NN}{k-nearest neighbours}
\newacronym{elev}{EV}{Elephant Voices}
\newacronym{elp}{ELP}{Elephant Listening Project}
\newacronym{dt}{DT}{decision tree}
\newacronym{rf}{RF}{random forest}
\newacronym{lr}{LR}{logistic regression}
\newacronym{mlp}{MLP}{multi-layer perceptron}
\newacronym{svm}{SVM}{support vector machine}
\newacronym{xgb}{XGB}{XGBoost}
\newacronym{dft}{DFT}{discrete Fourier transform}
\newacronym{lda}{LDA}{linear discriminant analysis}
\newacronym{lpc}{LPC}{linear predictive coding}
\newacronym{hmm}{HMM}{hidden Markov model}
\newacronym{pam}{PAM}{passive acoustic monitoring}
\newacronym{rnn}{RNN}{recurrent neural network}
\newacronym{map}{mAP}{mean average precision}
\newacronym{ap}{AP}{average precision}
\newacronym{pca}{PCA}{principal component analysis}
\newacronym{glm}{GLM}{generalised linear model}
\newacronym{resnet}{ResNet}{residual neural network}
\newacronym{iou}{IoU}{intersection over union}
\newacronym{beats}{BEATs}{bidirectional encoder representation from audio transformers}
\newacronym{iucn}{IUCN}{International Union for Conservation of Nature}
\newacronym{rbf}{RBF}{radial basis function}
\newacronym{umap}{UMAP}{uniform manifold approximation and projection}
\newacronym{gpu}{GPU}{graphics processing unit}
\newacronym{lmu}{LMU}{Legendre memory unit}
\newacronym{dec}{DEC}{deep embedded clustering}
\newacronym{nmi}{NMI}{normalised mutual information}
\newacronym{drl}{DRL}{disentangled representation learning}
\newacronym{vae}{VAE}{variational autoencoder}
\newacronym{lstm}{LSTM}{long short-term memory}
\newacronym{gru}{GRU}{gated recurrent unit}
\newacronym{tsne}{t-SNE}{t-distributed stochastic neighbour embedding}
\newacronym{gmm}{GMM}{Gaussian mixture model}
\newacronym{clap}{CLAP}{contrastive language-audio pretraining}
\newacronym{asr}{ASR}{automatic speech recognition}
\newacronym{mds}{MDS}{multidimensional scaling}
\newacronym{dtw}{DTW}{dynamic time warping}
\newacronym{aerd}{AERD}{automatic elephant rumble detection}
\newacronym{dct}{DCT}{discrete cosine transform}
\newacronym{HPC}{HPC}{High Performance Computing}

\newglossaryentry{librispeech-960h}{
	name={LibriSpeech ASR~(960h)},
	description={},
}

\newglossaryentry{beans}{
	name={BEANs},
	description={},
}

\newglossaryentry{beans20}{
	name={BEANs-20},
	description={},
}

\newglossaryentry{beans40}{
	name={BEANs-40},
	description={},
}

\newglossaryentry{elman}{
	name={Elman},
	description={},
}

\newglossaryentry{elman-rnn}{
	name={Elman RNN},
	description={},
}

\newglossaryentry{vggish}{
	name={VGGish},
	description={},
}

\newglossaryentry{vgg}{
	name={VGG},
	description={},
}

\newglossaryentry{birdnet}{
	name={BirdNET},
	description={},
}

\newglossaryentry{perchv1}{
	name={Perch~1.0},
	description={},
}

\newglossaryentry{perchv2}{
	name={Perch~2.0},
	description={},
}

\newglossaryentry{surfperch}{
	name={SurfPerch},
	description={},
}

\newglossaryentry{sierras}{
	name={Sierras},
	description={},
}

\newglossaryentry{psla}{
	name={PSLA},
	description={},
}

\newglossaryentry{yamnet}{
	name={YAMNet},
	description={},
}

\newglossaryentry{audiomae}{
	name={AudioMAE},
	description={},
}

\newglossaryentry{hubert}{
	name={HuBERT},
	description={},
}

\newglossaryentry{hubert-base}{
	name={HuBERT~(base)},
	description={},
}

\newglossaryentry{hubert-large}{
	name={HuBERT~(large)},
	description={},
}

\newglossaryentry{hubert-xlarge}{
	name={HuBERT~(xlarge)},
	description={},
}

\newglossaryentry{wav2vec2}{
	name={wav2vec2.0},
	description={},
}

\newglossaryentry{w2v2-base}{
	name={wav2vec2.0~(base)},
	description={},
}

\newglossaryentry{w2v2-large}{
	name={wav2vec2.0~(large)},
	description={},
}

\newglossaryentry{xlsr}{
	name={XLS-R},
	description={},
}

\newglossaryentry{multi-whale}{
	name={multi-species whale},
	description={},
}

\newglossaryentry{humpback}{
	name={humpback},
	description={},
}

\newglossaryentry{xeus}{
	name={XEUS},
	description={},
}

\newglossaryentry{wavlm}{
	name={WavLM},
	description={},
}

\newglossaryentry{aves}{
	name={AVES},
	description={},
}

\newglossaryentry{aves-core}{
	name={AVES-core},
	description={},
}

\newglossaryentry{aves-bio}{
	name={AVES-bio},
	description={},
}

\newglossaryentry{aves-nonbio}{
	name={AVES-nonbio},
	description={},
}

\newglossaryentry{aves-all}{
	name={AVES-all},
	description={},
}

\newglossaryentry{birdaves}{
	name={BirdAVES},
	description={},
}

\newglossaryentry{birdaves-biox-base}{
	name={BirdAVES-biox~(base)},
	description={},
}

\newglossaryentry{birdaves-biox-large}{
	name={BirdAVES-biox~(large)},
	description={},
}

\newglossaryentry{birdaves-bioxn-large}{
	name={BirdAVES-bioxn~(large)},
	description={},
}

\newglossaryentry{animal2vec}{
	name={animal2vec},
	description={},
}

\newglossaryentry{meerkat}{
	name={MeerKAT},
	description={},
}

\newglossaryentry{sincnet}{
	name={SincNet},
	description={},
}

\newglossaryentry{xenocanto}{
	name={Xeno-canto},
	description={},
}

\newglossaryentry{inaturalist}{
	name={iNaturalist},
	description={},
}

\newglossaryentry{efficientnet-b0}{
	name={EfficientNet-B0},
	description={},
}
\newglossaryentry{efficientnet-b1}{
	name={EfficientNet-B1},
	description={},
}
\newglossaryentry{efficientnet-b3}{
	name={EfficientNet-B3},
	description={},
}

\newglossaryentry{birdnetv2.4}{
	name={BirdNET~(ver.~2.4)},
	description={},
}

\newglossaryentry{fsd50k}{
	name={FSD50k},
	description={},
}

\newglossaryentry{audioset}{
	name={AudioSet},
	description={},
}

\newglossaryentry{vggsound}{
	name={VGGSound},
	description={},
}

\newglossaryentry{hdbscan}{
	name={HDBSCAN},
	description={},
}

\newglossaryentry{chatgpt}{
	name={ChatGPT},
	description={},
}

\newglossaryentry{ast-seq}{
	name={AST-seq},
	description={},
}

\newglossaryentry{ast-lab}{
	name={AST-cls},
	description={},
}

\newglossaryentry{lr-frame}{
	name={LR-frame},
	description={},
}

\newglossaryentry{librispeech}{
	name={LibriSpeech},
	description={},
}

\newglossaryentry{librivox}{
	name={LibriVox},
	description={},
}

\newglossaryentry{tierstimmenarchiv}{
	name={Tierstimmenarchiv},
	description={},
}

\newglossaryentry{birdclef}{
	name={BirdCLEF},
	description={},
}

\newglossaryentry{macaulay}{
	name={Macaulay Library of Natural Sounds},
	description={},
}

\usepackage{float}
\usepackage{caption}
\usepackage{subcaption}
\usepackage[xindy]{imakeidx}
\makeindex

\usepackage{enumitem}
\newlist{todolist}{itemize}{2}
\setlist[todolist]{label=$\square$}
\usepackage{pifont}
\usepackage{bbding}

\usepackage[most]{tcolorbox}
\newtcolorbox{notebox}{parbox=false, colback=gray!15!white, colframe=gray!75!black, fonttitle={\scriptsize\scshape}, enhanced, attach boxed title to bottom right={yshift=3mm,xshift=-4mm}, title={Note}, }

\usepackage{doi}

\usepackage{siunitx}
\usepackage[title]{appendix}
\usepackage{etoolbox}
\AtBeginEnvironment{appendices}{\crefalias{section}{appendix}} \usepackage[capitalise, nameinlink]{cleveref}

\newenvironment{resultstable}{
  \sisetup{
    detect-all,
    mode=math,
    table-alignment-mode = format,
    table-format = 1.3,
    table-number-alignment = left,
    separate-uncertainty = false,
    round-precision = 4,           }
  \renewrobustcmd{\bfseries}{\fontseries{b}\selectfont}
  \renewrobustcmd{\boldmath}{}
\newrobustcmd{\B}{\bfseries}
  \newrobustcmd{\U}[1]{\uline{\num{##1}}}
  \newcommand{\embedding}[3][5]{\multirow{##1}{*}{##2} & \multirow{##1}{*}{##3}}
  \newcommand{\resultsheading}{
&                &        &  & \multicolumn{2}{c}{EV}    &  & \multicolumn{2}{c}{LDC}   \\* \cmidrule(lr){5-6} \cmidrule(lr){8-9}
    Embedding & Dim. & Class. &  &         AUC &         mAP &  &         AUC &         mAP \\*
}
\newcommand{\subheading}[1]{
\addlinespace[0.5em]
    \midrule
    \multicolumn{9}{c}{\textit{##1}} \\
    \midrule
}
  \newcommand{\subheadinglabel}[1]{
  \begin{subtable}{\textwidth}
	\phantomcaption\label{##1}
  \end{subtable}
  }
\newcommand{\newembed}{
    \addlinespace[0.25em]
    \cmidrule[0.25pt](l){5-9}
\addlinespace[0.25em]
  }
}{}
\DeclareSIUnit{\unitless}{\relax}

\usepackage[normalem]{ulem}

\robustify\bfseries
\robustify\uline
\robustify\tnote

\author{
    \orcidlinki{Christiaan M. Geldenhuys}{0000-0003-0691-0235} \\
	Department of Electrical and Electronic Engineering\\
	University of Stellenbosch\\
	\href{mailto:cmgeldenhuys@sun.ac.za}{\texttt{cmgeldenhuys@sun.ac.za}} \\
\And
	\orcidlinki{Thomas R. Niesler}{0000-0002-7341-1017} \\
	Department of Electrical and Electronic Engineering\\
	University of Stellenbosch\\
	\href{mailto:trn@sun.ac.za}{\texttt{trn@sun.ac.za}} \\
}

\title{From Birdsong to Rumbles}
\renewcommand{\headeright}{Preprint}
\renewcommand{\undertitle}{Classifying Elephant Calls with Out-of-Species Embeddings}
\renewcommand{\shorttitle}{\textsc{\scriptsize From Birdsong to Rumbles: Classifying Elephant Calls with Out-of-Species Embeddings}}
\renewcommand{\subsubsubsection}[1]{\textbf{#1}.~}

\hypersetup{
pdftitle={From Birdsong to Rumbles: Classifying Elephant Calls with Out-of-Species Embeddings},
pdfsubject={Bioacoustics, Elephant Vocalisation Classification},
pdfauthor={Christiaan M.~Geldenhuys, Thomas R.~Niesler},
pdfkeywords={elephant vocalisation, bioacoustics, pretrained embeddings, transfer learning, call classification},
}

\begin{document}
\maketitle
\begin{abstract}
	We show that pretrained acoustic embeddings classify elephant vocalisations at a level approaching that of end-to-end supervised neural networks, without any fine-tuning of the embedding model.
	This result is of practical importance because annotated bioacoustic data are scarce and costly to obtain, leaving conventional supervised approaches prone to overfitting and to poor generalisation under domain shift.
	A broad range of embedding models drawn from general audio, speech, and bioacoustic domains is evaluated, all of which are either \textit{out-of-domain}~(containing no bioacoustic data) or \textit{out-of-species}~(containing no elephant call data).
	The embedding networks themselves remain fixed; only the lightweight downstream classifiers, which include a linear model and several small neural networks, are trained.
	Among the models considered, \gls{perchv2} achieves the best cross-validated classification performance, attaining \acsp{auc} of \num{0.849} on African bush elephant~(\textit{Loxodonta africana}) calls and \num{0.936} on Asian elephant~(\textit{Elephas maximus}) calls, with \gls{perchv1} close behind.
	The best-performing system is within \qty{2.2}{\percent} of an end-to-end supervised elephant call classification system.
	A layerwise analysis of pretrained transformer encoders, considered as embedding models, shows that intermediate representations outperform final-layer outputs.
	The second layer of both \gls{wav2vec2} and \gls{hubert} encodes sufficient information for effective elephant call classification; truncation at this layer therefore preserves classification performance whilst retaining only approximately \qty{10}{\percent} of the parameters of the full network.
	Such compact embedding networks are well suited to on-device processing where computational resources are limited.
\end{abstract}

\section{Introduction}
\label{sec:intro}

The African bush elephant (\textit{Loxodonta africana}) and African forest elephant (\textit{L. cyclotis}) are currently classified as endangered and critically endangered species, respectively~\autocite{iucn2020loxodontaafricana, iucn2020loxodontacyclotis}.
Population decline in both species is driven principally by habitat destruction and illegal hunting.
The Asian elephant (\textit{Elephas maximus}) is similarly designated as endangered by the IUCN.
Beyond ivory poaching, this species is confronted by serious challenges from the human-elephant conflict~\autocite{iucn2019elephasmaximus}.

Automated classification of elephant vocalisations offers a promising opportunity to advance the understanding of elephant behavioural patterns, their population distribution, and the conservation requirements~\autocite{zeppelzauer2015eledetsystem, keen2017elepcnn}.
The growing deployment of \acp{aru} in protected areas and conservation facilities has produced archives of continuous \ac{pam} audio whose scale far exceeds what can be reviewed manually, making automated analysis a prerequisite for turning such recordings into ecological insight.
Such systems can inform wildlife management strategies, act as an early detection mechanism for conservation threats including poaching activities, and yield behavioural data of value to bioacoustic research and ecological studies.

Acoustic embedding models are typically pretrained on large collections of audio data, commonly using self-supervised learning, which does not require data labels.
Although these pretraining datasets do not necessarily align with the intended target domain, the learned high-dimensional vector representations are hypothesised to encode information that generalises beyond the pretraining domain, a setting we refer to as \textit{out-of-domain}.
The same principle extends to embedding models pretrained on other fauna; we refer to such models as \textit{out-of-species} when their training corpora contain no vocalisations of the target species.
We apply this criterion strictly: a model whose pretraining data is known or reasonably suspected to include recordings of the target species is not considered out-of-species, even if such recordings constitute only a small fraction of the corpus.

Classification based on pretrained embeddings provides a data-efficient means of developing complex classifiers that can perform well despite limited training data.
Such embedding-based approaches have shown considerable promise in bioacoustics, achieving strong performance for bird species identification~\autocite{google2023perch} and marine mammal call classification~\autocite{zhong2021siamesewhale}, but have not yet been applied to elephant vocalisation classification.

In this work, we consider embedding models that have been pretrained on general audio or on human speech, as well as on avian, terrestrial, and marine bioacoustic recordings.
Critically, these embedding models remain fixed and are not subject to any further training throughout our experiments.
This follows the standard evaluation paradigm for self-supervised audio models~\autocite{Yang2021SUPERB}, in which fixed representations are paired with lightweight downstream classifiers.
The approach is particularly appropriate in the work we present here, given the substantial mismatch in scale between our target datasets and the embedding models, which contain tens to hundreds of millions of parameters.
Fine-tuning such architectures on limited data has been shown to lead to overfitting, even for in-domain downstream tasks~\autocite{chen2022wavlm}.
The extracted embeddings are therefore used as fixed feature inputs to train several smaller neural classifiers in a supervised classification framework.
The classifiers considered are \acf{lr}, \acfp{mlp}, Elman \acfp{rnn}, \acfp{gru}, and \acf{lstm} networks.
We compare the performance of these classifiers with that of a baseline \acf{ast} model trained end-to-end for elephant call classification~\autocite{geldenhuys2024aerd}, in which all model parameters were directly optimised on the target data.
Each classifier is evaluated against the expert annotations using \ac{auc} and \ac{map} as performance indicators.

For the best-performing transformer embedding models, we perform an intermediate layerwise analysis.
Intermediate transformer layers have been shown to model hierarchical aspects of human speech: earlier layers encode phonetics, later layers encode words, and the final layers encode intent~\autocite{pasad2021layerwise}.
Elephant calls have also been shown to exhibit structured communication mechanisms that convey emotion and coordinate herd behaviour~\autocite{Soltis2005ElephantVocalShortRange,Poole2005VocalLearning,stoeger2012lda}.
Specific transformer layers that encode such structure may therefore yield embeddings better suited to call classification.
Such analysis also carries a direct practical motivation: if intermediate layers suffice for classification, only a fraction of the pretrained model need be retained at inference time, reducing the compute footprint to a level compatible with on-device deployment in remote conservation settings.

\subsection{Contributions}
To the best of our knowledge, the work presented in this paper makes the following contributions:
\begin{itemize}
	\item
	      We present the first evaluation of elephant call classification using embeddings obtained from a broad range of pretrained neural networks.
	      Using two separate datasets, we show that this approach can achieve classification performance that approaches that of end-to-end supervised models in a low-resource setting, without any fine-tuning of the embedding models.
	\item
	      We present the first layerwise analysis of intermediate transformer representations for elephant call classification.
\end{itemize}

\section{Background}
We provide a brief overview of elephant call classification and acoustic embedding models.

\subsection{Elephant call classification}
The study of wildlife acoustics has emerged as a critical area of research at the confluence of ecology, conservation biology, and machine learning.
Elephants are known to exhibit complex social structures, inter-herd communication networks, and remarkable cognitive abilities, making understanding their vocal behaviour key to furthering conservation efforts and socio-ecological studies~\autocite{soltis2010vocal,zeppelzauer2015autoclass}.
Elephant rumbles, which are low-frequency vocalisations, convey information such as age, sex, reproductive status, and emotional state~\autocite{poole1988social, poole1994sex}.

Although automated elephant call classification has been attempted~\autocite{clemins2003elepspeech,zeppelzauer2015autoclass}, data scarcity has limited the use of more advanced neural approaches, which have consequently not yet delivered substantial performance gains~\autocite{bjorck2019elepcnn,geldenhuys2024aerd}.
Gathering and annotating further recordings is expensive, and the labelling itself requires trained volunteers alongside domain experts in bioacoustics or ecology to identify and label vocalisations correctly.
Even under strict annotation conventions, labels may be noisy due to disagreements in annotated start and end times, calls obscured by interfering sources, overlapping vocalisations, variability in recorded call structure, or inter-annotator bias.

Given these constraints, an attractive alternative to training specialised models from scratch is to reuse the representations learned by neural networks pretrained on large audio corpora collected for other purposes.
Such corpora, which span general audio, human speech, and bioacoustic recordings of other species, are orders of magnitude larger than any available elephant call dataset, and the resulting representations might capture acoustic structure that transfers to elephant vocalisations despite the domain shift.

\subsection{Acoustic embedding models}
Acoustic embedding models are deep neural network models that seek to represent the information in an audio signal as a high-dimensional numerical vector with fixed dimensionality.
When applied to speech, for example, phonological and semantic information has been found to be encoded in the orientation of the resulting vector~\autocite{Bengio2013RepresentationLearning, pasad2021layerwise}.
We refer to these representative vectors as acoustic embeddings, or simply as embeddings.
Embedding models are typically trained on very large but unlabelled audio datasets using self-supervised or unsupervised machine learning techniques.
During training, the embedding model is tasked with discovering the underlying patterns that represent the data without the aid of human annotations.
This can be achieved using approaches such as contrastive learning or masked token prediction~\autocite{Saeed2021COLA,baevski2020wav2vec2}.
Because the data is unlabelled, the risk of human bias in the labels is reduced.
However, these approaches require substantially more data and computation than supervised alternatives.

\section{Literature Review}
The application of deep learning to bioacoustics has grown substantially in recent years, with \textcite{stowell2022biodeep} providing a comprehensive review.
Here, we focus on work most relevant to the use of pretrained embedding models for bioacoustic classification, as well as on existing approaches to elephant call classification.

\subsection{Pretrained embeddings in bioacoustics}
Early work in marine bioacoustics demonstrated that learned acoustic representations can encode species- and population-level information.
\Textcite{zhong2021siamesewhale} employed embeddings obtained from a Siamese neural network to detect, classify, and count the calls of four blue whale populations from \qty{350}{hours} of manually annotated hydrophone recordings.
A \ac{tsne} visualisation of the embedding space revealed clear cluster separation between populations, indicating that the model encodes population-specific acoustic signatures.
Similarly, \Textcite{ozanich2021whalecluster} applied \ac{dec} to embeddings obtained from models trained on unlabelled coral reef recordings, achieving clear separation between whale song units and fish pulse calls.
These foundational studies established the viability of embedding-based approaches for wildlife acoustic classification.

Subsequent work examined whether pretrained embeddings generalise across broader taxonomic groups.
\Textcite{sethi2022vggraretaxa} performed rare species identification in a tropical forest environment using embeddings obtained from a pretrained \gls{vggish} model on a weakly-labelled dataset, demonstrating that soundscape-level analysis offers a data-efficient alternative to species-specific call detection.
Most comprehensively, \Textcite{google2023perch} evaluated embeddings from six \ac{cnn} architectures and one transformer architecture across several animal vocalisation datasets spanning birds, bats, marine mammals, and amphibians.
Notably, none of the embedding models had been explicitly trained on any of the taxa present in the evaluation sets.
The authors found that models trained on bird calls~(\gls{birdnet} and \gls{perchv1}) outperformed all other architectures, including the transformer model, suggesting that training domain influences transfer effectiveness in a multi-taxa setting.

Beyond species-level classification, researchers have investigated whether embeddings capture finer-grained acoustic variation.
\Textcite{best2023vocalcluster} obtained embeddings using a \ac{cnn} auto-encoder and applied unsupervised clustering to perform acoustic unit discovery, finding substantial agreement between automatically discovered units and expert-annotated vocalisation repertoires.
\Textcite{mcginn2024avianid} showed that \gls{birdnet} embeddings can distinguish between adult and juvenile Great Grey Owl vocalisations and identify different call types of the Great Spotted Woodpecker through unsupervised clustering.
\Textcite{bravo2024bioemb} extended these findings, demonstrating that embedding clusters represent not only different species at a global level but also different vocalisations within a single species.
Collectively, these studies suggest that pretrained embeddings encode meaningful acoustic structure at multiple taxonomic and behavioural levels.

Concurrently, researchers have investigated whether models trained on human speech can serve as effective acoustic feature extractors for bioacoustic tasks.
Recent work by \Textcite{Cauzinille2025SpeechTransferBio} provides compelling evidence for the transferability of self-supervised speech models to bioacoustic classification.
The authors benchmarked three speech models (\gls{hubert}, \gls{wavlm}, and \gls{xeus}) on eleven datasets spanning birds, mammals, and insects, demonstrating competitive performance with fine-tuned bioacoustic models whilst substantially outperforming \ac{cnn} baselines.
Notably, phylogenetic proximity to humans did not influence transfer performance, and the authors found that simple linear classifiers performed comparably to more complex recurrent classifiers, which they concluded was likely due to overfitting.

\subsection{Elephant call classification}
Automatic elephant vocalisation classification has received considerable attention in the bioacoustic literature, though existing computational approaches differ substantially from the embedding-based methods described above.
\Textcite{clemins2003elepspeech,clemins2005elephmm} conducted the first application of speech processing techniques to elephant call classification, employing \acp{mfcc} with a manually adjusted mel-scale to account for the infrasonic nature of elephant calls.
Using a \ac{hmm} classifier, the authors achieved \qty{79.74}{\percent} accuracy on a five-class classification task.
\Textcite{stoeger2012lda} applied \ac{lda}, \ac{svm}, and \ac{knn} classifiers to features obtained from \ac{lpc}-smoothed spectrograms, achieving \qty{99}{\percent} accuracy in distinguishing oral from nasal rumbles.

For call detection, \Textcite{zeppelzauer2015autoclass} were the first to apply automatic classification to free-roaming elephants, using an \ac{svm} classifier with Greenwood cepstrum features.
\Textcite{keen2017elepcnn} improved upon this with handcrafted two-dimensional convolutional kernels, achieving \qty{87.2}{\percent} sensitivity and \qty{91.0}{\percent} specificity.
More recently, deep learning architectures such as \acp{cnn} and CNN-LSTM hybrids have been applied to elephant call detection with promising results~\autocite{bjorck2019elepcnn,leonid2022elepcnn}.
\Textcite{geldenhuys2024aerd} demonstrated that fine-tuning a pretrained \ac{ast} model yields state-of-the-art performance for both call activity detection and call classification, achieving an average precision of 0.962 for binary detection and an \ac{auc} of 0.957 for five-class call classification.
Later, \textcite{Dewmini2025ElephantCallerNet} proposed \textit{ElephantCallerNet}, a lightweight raw-waveform \ac{cnn} inspired by the \textit{ACDNet} architecture and designed for edge deployment, reporting \qty{89}{\percent} test accuracy on three-class Asian elephant caller-type classification.

Most closely related to the present work, \Textcite{Pickering2025ForestElephantEmbeddings} applied four pretrained \ac{cnn} embedding models, comprising two general-purpose~(\gls{vggish} and \gls{yamnet}) and two avian bioacoustic~(\gls{birdnet} and \gls{perchv1}) networks, to \num{1254} manually annotated African forest elephant~(\textit{Loxodonta cyclotis}) vocalisations recorded at Dzanga-Bai, Central African Republic.
A \ac{rf} classifier trained on a \ac{umap} of the embeddings to discriminate the three call types in the corpus (rumble, roar and trumpet) yielded macro-average accuracies of \num{0.85} for \gls{perchv1}, \num{0.84} for \gls{birdnet} and \gls{vggish}, and \num{0.81} for \gls{yamnet}.
Unsupervised affinity-propagation clustering of the same \ac{umap} projections, restricted to rumbles, recovered five to six sub-types per model, which the authors interpreted in terms of the demographic and behavioural composition of the constituent calls.
The resulting groupings were broadly consistent across the four embedding architectures, and a complementary \ac{glm} analysis further identified age and behavioural context as significant predictors of acoustic variation~($p<0.001$).
Together these findings suggest that pretrained embeddings are able to capture behavioural and demographic structure within single call types.

Despite this progress, existing approaches to elephant call classification and detection have predominantly relied on either handcrafted acoustic features paired with traditional classifiers, or deep neural networks trained or fine-tuned specifically on elephant data.
With the exception of the contribution by \textcite{Pickering2025ForestElephantEmbeddings}, the application of pretrained embedding models to elephant vocalisation classification remains unexplored.
No prior work has systematically evaluated such embeddings across multiple model families, downstream probe configurations, intermediate layer representations, or elephant species.We address this gap by evaluating a broad collection of embedding models, including previously unexplored transformer and self-supervised speech architectures, each paired with a range of downstream classifiers.
Every embedding and classifier pair is compared across two datasets, covering African and Asian elephant vocalisations respectively.
Together these corpora comprise \num{4945} annotated calls, affording a broad empirical basis for assessing embedding performance.
For a selection of the top-performing transformer architectures, a layerwise analysis is performed to determine whether particular layers are especially effective at encoding elephant vocalisations, as has been observed for human speech.

\section{Datasets}
\label{sec:data}
We employ two datasets which we will refer to as the \ac{elev} dataset and the \ac{ldc} dataset, respectively.
Both datasets are evaluated individually; in other words, all models will be trained and tested separately for each respective dataset.
This is done due to differences in species, recording setups and size of the datasets.
\Cref{tab:data:summary} provides key information for each dataset.

\begin{table}
	\centering
	\caption{
		Summary of \acf{elev} and \acf{ldc} elephant vocalisations datasets used for experimentation, adapted from \textcite{geldenhuys2024aerd}.
		All recording statistics are provided on the original long form recordings, before any division.
}
	\begin{tabular}{@{}rcc@{}}
		\toprule
		Dataset                            & \textit{\acs{elev}}         & \textit{\acs{ldc}}              \\ \midrule
		Authors                            & \textcite{poole2021elev}    & \textcite{ldc2010asianelevoc}   \\
		Elephant species                   & \textit{Loxodonta africana} & \textit{Elephas maximus}        \\
		Recording environment              & Handheld field recordings   & Handheld field recordings       \\
		Recording equipment                & ARES-BB Nagra               & Fostex FR-2                     \\
		Microphone                         & Not specified               & Earthworks QTC50                \\
		Number of call types               & 33                          & 16                              \\
		Detail of annotation               & File-level                  & Within \qty{100}{\milli\second} \\
		Total annotated duration           & \qty{36}{minutes}           & \qty{5.4}{hours}                \\
		Number of vocalisations (segments) & 514                         & 4433                            \\
		Number of recordings               & 226                         & 1577                            \\
		Average length                     & \qty{14.75}{\second}        & \qty{131.03}{\second}           \\
		Min.                               & \qty{0.49}{\second}         & \qty{1.25}{\second}             \\
		Max.                               & \qty{296.52}{\second}       & \qty{3889.38}{\second}          \\
		Std. dev.                          & \qty{31.48}{\second}        & \qty{177.02}{\second}           \\
		Sampling rate                      & \qty{44.1}{\kilo\hertz}     & \qty{16}{\kilo\hertz}           \\
		Bit depth                          & 16-bit                      & 24-bit                          \\
		Number of channels                 & 2                           & 1                               \\
		Low frequency cut-off              & Not specified               & \qty{3}{\hertz}                 \\
		Cross-validation folds             & 5-fold                      & 10-fold                         \\ \bottomrule
	\end{tabular}
	\label{tab:data:summary}
\end{table}

The \ac{elev} dataset consists of 60~minutes of field recordings made by \citeauthor{poole2021elev} at the non-profit organisation Elephant Voices~\autocite{poole2021elev}.
The recordings consist of the vocalisations of free-roaming African bush elephants~(\textit{Loxodonta africana}) and include 33~types of calls and subcalls.
Annotations are provided at a recording level, meaning that no time locality is provided.
Some recordings are long and may include multiple calls.
In the work presented here, these have been divided into shorter recordings~(segments) containing only a single call per recording~(\cref{sec:segmentation}), and where the start and end times of each call segment have been manually annotated by the authors.
The dataset has been partitioned into five disjoint subsets, used for nested K-fold cross-validation~\autocite{stone1974crossval} in our experiments.

The \ac{ldc} dataset consists of 30~hours of annotated field recordings made by \citeauthor{ldc2010asianelevoc} and hosted by the \ac{ldc}~\autocite{ldc2010asianelevoc}.
The recordings contain the vocalisations of free-roaming Asian elephants~(\textit{Elephas maximus}).
The dataset has been annotated according to the call taxonomy presented in \textcite{ldc2010asianelevoc} and contains 16~distinct call types.
Annotations are provided in the form of start- and end-times within a long form recording.
As before, these long form recordings are divided into shorter recordings~(segments) containing only a single call, with a collar of \qty{250}{\milli\second} added to the start and end of the recording.
The dataset has been partitioned into ten disjoint subsets, used for nested K-fold cross-validation~\autocite{stone1974crossval} in our experiments.

\section{Experimental Setup}
\begin{figure}[hp]
	\centering
	\begin{subfigure}{1.0\linewidth}
		\centering
		\caption{
			Audio segmentation process.
		}
		\label{fig:structure:segment}
		\includegraphics[page=1]{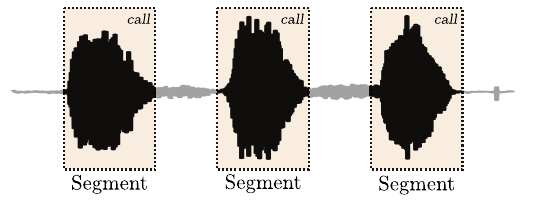}
	\end{subfigure}
	\begin{subfigure}{1.0\linewidth}
		\centering
		\caption{
			Embedding extraction process.
		}
		\label{fig:structure:extract}
		\includegraphics[page=2]{model/experiment_structure}
	\end{subfigure}
	\begin{subfigure}{1.0\linewidth}
		\centering
		\caption{
			Classification using feed-forward models.
		}
		\label{fig:structure:flat}
		\includegraphics[page=3]{model/experiment_structure}
	\end{subfigure}
	\begin{subfigure}{1.0\linewidth}
		\centering
		\caption{
			Classification using recurrent architectures.
		}
		\label{fig:structure:recurrent}
		\includegraphics[page=4]{model/experiment_structure}
	\end{subfigure}
	\caption{
		Illustration of the experimental structure used throughout the manuscript.
		(\subref{fig:structure:segment})~First, audio signals are segmented~(\cref{sec:segmentation}) into shorter intervals, each containing a single call; (\subref{fig:structure:extract})~these audio segments are presented to an embedding model~(\cref{sec:emb-model}) from which a sequence of embedding vectors is obtained, one per time frame, representing the information captured within the audio signal.
		These embeddings are then used to perform the final classification~(\cref{sec:shallow-classifier}) using either: (\subref{fig:structure:flat})~\acf{lr} or \acf{mlp}, for which the sequence of embeddings for a given segment is averaged before classification; or (\subref{fig:structure:recurrent}) an Elman \acf{rnn}, a \acf{lstm}, or a \acf{gru}, which recurrently consume the sequence of embeddings and produce the final classification from the final hidden state.
	}
	\label{fig:structure}
\end{figure}

This section describes the experimental procedure used to perform elephant call classification with pretrained embeddings.
First, we describe the audio segmentation process, in which a long audio recording is divided into smaller segments.
This is followed by a description of the models used to obtain the embeddings and the configuration of the various classification models that use these embeddings.
Importantly, all embedding models remain fixed throughout; only the classification model weights are optimised.
Finally, the metrics used for performance evaluation are described.
\Cref{fig:structure} illustrates the structure of the experiments.

\subsection{Audio segmentation}
\label{sec:segmentation}
Segmentation refers to the process of isolating the intervals in a long audio recording that correspond to elephant vocalisations.
Therefore, during segmentation the start and end times of each call are determined, a process sometimes also referred to as endpointing.
Endpointing can be achieved manually or automatically.
We do not consider the endpointing procedure here.
Instead, we assume that endpoints are known and use the segments that are provided in the human annotations.
Although vocalisations may overlap temporally in the dataset, each segment is defined by a single endpointed call and we therefore formulate classification as a multi-class problem.
The training target for each segment is determined by the vocalisation present at the temporal centre of that segment.
At evaluation time, a segment may contain additional overlapping vocalisations and a classification is therefore deemed correct if it corresponds to any of the human-annotated labels within the segment.

\subsection{Embedding models}
\label{sec:emb-model}
\label{sec:emb}
The models described in this section all produce a sequence of embeddings.
Unless otherwise stated, one embedding is produced per frame.
In the case of models using spectral features, these frames are those used by the \acf{stft}.
In the case of end-to-end models such as \gls{wav2vec2}, these frames correspond to the length of the input kernels in the first \ac{cnn} layer, and the rate at which frames are produced is determined by the inter-kernel stride.
Each embedding takes the form of a high-dimensional numerical vector for the corresponding input audio frame, as illustrated in \cref{fig:structure:extract}.
The purpose of the embeddings is to represent the information contained in the input audio data in a high dimensional space.
This mapping is optimised to represent the sounds that were present in the dataset that was used for training the embedding model.
From our point of view, these embeddings would ideally be representative of the information present within the audio signal, such as animal identity, vocalisation information, and acoustic environment.
One of our explicit aims is therefore to establish whether the embeddings computed by out-of-domain or out-of-species models are generalisable to elephant call classification.

The section is divided into four subsections: \textit{baseline}, \textit{general acoustics}, \textit{speech}, and \textit{bioacoustics}.
Each of these represents a different embedding model architecture, a different training objective, or a different pretraining dataset.

\subsubsection{Baseline embeddings}
\label{sec:emb:baseline}
In this subsection, we describe classification methods we will use as baselines.
\Ac{mfcc} and \gls{beans} are spectral feature extraction methods that we will treat as embedding models, whilst \gls{aerd} is an end-to-end supervised classification model.

\subsubsubsection{MFCCs}
The first baseline model uses the mean of the \ac{mfcc} vectors computed within the audio segment as the embedding.
\Acp{mfcc} are computed following the standard procedure described by \textcite{Davis1980MelConfig}, using a \qty{25}{\milli\second} frame length with a \qty{10}{\milli\second} stride and a Hanning window function.
A 1024-point \ac{fft} is used to obtain the power spectrum, and a mel-scale filterbank with 128 triangular filters is applied.
The lowest 40 cepstral coefficients are retained.

\subsubsubsection{BEANs}
\Textcite{hagiwara2022beans} propose \gls{beans}, a widely used bioacoustic benchmark model for vocalisation classification.
The model follows the same process set out for \ac{mfcc} features above.
However, instead of only computing the mean over the segment, the standard deviation, minimum, and maximum are also computed for each vector dimension.
These four aggregated feature vectors are concatenated to form a single vector to act as the embedding.
The \gls{beans} embedding uses a \qty{50}{\milli\second} frame length and a \qty{10}{\milli\second} stride between frames.
We evaluate two variants of the \gls{beans} embeddings: \gls{beans20}, which retains the lower 20~\ac{mfcc} coefficients, and \gls{beans40}, which retains the lower 40~\acp{mfcc}.

\subsubsubsection{AERD}
The final benchmark we consider is a \ac{vit}[-based] model operating on mel spectrogram input features, trained using the \gls{beats} self-supervised training regime~(\cref{sec:emb:beats}) and subsequently fine-tuned end-to-end for elephant call classification~\autocite{geldenhuys2024aerd}.
Although \textcite{geldenhuys2024aerd} evaluate several model configurations, we consider only their best-performing variant, labelled \textit{AST-seq}, and refer to it as \textit{\gls{aerd}} throughout.
No embedding is obtained from this model.
Instead, it is evaluated directly on the classification task.
Thus, \gls{aerd} acts as a performant, end-to-end supervised benchmark classification model in which all parameters were optimised on the target elephant call data.

\subsubsection{General audio embedding models}
\label{sec:emb:general}
This subsection describes two models which have been trained on general audio datasets and are therefore not specific to bioacoustics or speech: \gls{vggish} is a \ac{cnn} model, whilst \gls{beats} is a transformer model.
Both models take spectral representations as input, as opposed to being directly applied to the audio waveform.

\subsubsubsection{BEATs}
\label{sec:emb:beats}
\Textcite{chen2022beats} present \gls{beats}, in which a \acf{vit} encoder is trained on mel spectrogram features using an iterative self-distillation procedure: quantised codebook embeddings are distilled into the encoder through masked token prediction~\autocite{devlin2019bert}, after which the trained encoder is used to retrain the codebook, and this process is repeated.
The model is pretrained on \gls{audioset}~\autocite{gemmeke2017audioset}, a weakly labelled dataset comprising \qty{5800}{hours}~(approximately) of \qty{10}{\second} audio clips spanning \num{632} sound event classes.
Representations are obtained from the final transformer layer, omitting the linear classification layer.
These form a two-dimensional grid indexed by time step and spectral~(frequency-patch) position, each contextualised by the self-attention mechanism over the full input.
We consider three strategies for aggregating this grid into an embedding sequence.
The \gls{beats}~(time+spec) variant flattens the grid into a single sequence, retaining both axes.
The \gls{beats}~(time) variant averages over the spectral axis, yielding one embedding per time step, whilst the \gls{beats}~(spec) variant averages over the time axis, yielding one embedding per spectral patch.
Because averaging over time removes the sequential structure, recurrent classifiers are not applicable to the \gls{beats}~(spec) variant, and thus omitted from our experiments.

\subsubsubsection{VGGish}
\Gls{vggish} is a \ac{cnn} architecture developed by \textcite{hershey2017cnnaudio} for general audio processing tasks, trained on \gls{audioset}.
Inspired by the \gls{vgg} image classification architecture~\autocite{simonyan2015vgg}, \gls{vggish} operates on spectrogram representations of the input audio, allowing image processing techniques to be applied.
Through a series of convolutional and pooling layers, the model extracts hierarchical auditory features, which are then processed by a \ac{mlp} to obtain the final embeddings.

\subsubsection{Speech embedding models}
\label{sec:emb:speech}
This section describes three embedding models that are currently effective for speech recognition and are trained using self-supervision on recordings of monolingual and multilingual speech.

\subsubsubsection{\Gls{wav2vec2}}
\Textcite{baevski2020wav2vec2} introduces a training technique applied to a transformer encoder architecture, which is tasked with determining the masked quantised codebook embedding from raw audio whilst contrastively separating positive and negative pairs in a latent space.
First, a \ac{cnn} model is used to obtain a series of latent speech features~$\mathcal{Z}$.
These features are quantised using a straight-through codebook embedding to obtain the quantised representation~$\mathcal{Q}$ of the speech vectors.
From the latent speech features~$\mathcal{Z}$, a transformer encoder model produces a one-to-one contextualised representation~$\mathcal{C}$ for each latent vector, where contextualisation is accomplished through the self-attention mechanism.
A masked contrastive loss term is computed between the quantised representation~$\mathcal{Q}$ and the masked contextualised representations~$\mathcal{C}$.
\Textcite{baevski2020wav2vec2} presents several model size variants with increasing numbers of transformer layers, heads, and embedding dimensions: \textit{base}, \textit{large}, and \textit{extra-large}.
Our main experiments focus on the base variant of \gls{wav2vec2}, trained on the \gls{librispeech-960h} corpus~\autocite{openslr2015librispeech}, chosen because its parameter count is comparable to that of the other embedding models we evaluate.
Embeddings are obtained from the final transformer layer.
The large and extra-large variants are included for completeness; their results appear in \cref{app:results:class}.

\subsubsubsection{\Gls{xlsr}}
\Textcite{conneua2020xlsr} trained a \gls{wav2vec2} model using a larger multilingual dataset containing noisier data, wider speaker diversity, and different recording environments.
The dataset of \gls{xlsr} contains roughly \qty{436000}{hours} of speech across \num{128} languages, compared to \gls{wav2vec2} which was pretrained on roughly \qty{56000}{hours} of unlabelled English speech from \gls{librivox}~\autocite{mcguire2005librivox} and fine-tuned on \qty{960}{hours} of labelled English speech from \gls{librispeech}.
Again, the embedding is obtained from the final transformer layer.

\subsubsubsection{\Gls{hubert}}
\Textcite{hsu2021hubert} extend the work of \textcite{baevski2020wav2vec2} by replacing the masked contrastive loss function and quantised codebook embeddings with an unsupervised acoustic discovery unit clustering algorithm.
During training, \gls{hubert} is tasked with correctly determining the acoustic cluster, computed from the speech features, belonging to the contextualised representation.
This, in conjunction with a teacher-student distillation technique, showed improved performance in \ac{asr} tasks compared to \gls{wav2vec2}.
As with \gls{wav2vec2}, our main experiments use the base variant of \gls{hubert}, trained on the \gls{librispeech-960h} corpus.
Embeddings are obtained from the final transformer layer, omitting the acoustic cluster identification layer.
The large and extra-large variants are included for completeness; their results appear in \cref{app:results:class}.

\subsubsection{Bioacoustic embeddings}
\label{sec:emb:bio}
This section describes seven embedding models trained specifically on bioacoustic data, both terrestrial and marine.
With the exception of \gls{aves}, \gls{birdaves}, and \gls{animal2vec}, all models in this section are \ac{cnn} models.

\subsubsubsection{BirdNET}
First introduced by \textcite{kahl2021birdnet}, \gls{birdnet} is a \ac{resnet}, developed for automated identification of bird species from audio recordings.
The model has evolved since it was first announced.
At the time of writing, the newest version was \gls{birdnet} version 2.4, which was trained to recognise over \num{6000} bird species globally by identifying species-specific calls in audio recordings.
The training data comprised \num{226078} audio recordings totalling over \num{3978} hours, sourced primarily from \gls{xenocanto}~\autocite{vellinga2015xenocanto}, the \gls{macaulay}~\autocite{scholes2015macaulay}, and proprietary recordings from the BirdNET mobile application.
The dataset consisted primarily of focal-follow recordings, with a high signal-to-noise ratio and recordings typically containing a single, clearly audible species -- rather than soundscape recordings as captured by \acp{aru}.
However, the set also includes a number~(roughly \qty{10}{\percent}) of soundscape recordings from \gls{birdclef}~\autocite{Khal2024BirdCLEFOverview}.
Each recording included extensive metadata with quality scores, which were used to select high-quality samples for training.
Manual verification of species labels was not practical given the dataset size.
A maximum of 500 recordings per species was retrieved from the available data, and species with fewer than 10 recordings were excluded from the final training set.
\Gls{birdnetv2.4} is based on the \gls{efficientnet-b0} architecture, with an embedding dimensionality that has been reduced from \num{1280} to \num{1024}.
The model computes embeddings using a trainable mel-spectrogram implementation with 64 mel-frequency bands~\autocite{Sprengel2016SpecErosion}.
Embeddings are obtained from the final pooling layer, omitting the final linear classification layer.

\subsubsubsection{Perch 1.0}
The \gls{perchv1}~\autocite{google2023perch} model relies on the larger \gls{efficientnet-b1} \ac{cnn} architecture, with no alteration to model architecture.
The model is trained in a supervised fashion, exclusively on bird vocalisation data from \gls{xenocanto}.
The training data consisted of weakly labelled recordings (a single label per recording), and an activity detector was used to select training windows from each file.
Unlabelled recordings were automatically labelled using proprietary commercial machine learning systems (Google AutoML project).
While \gls{birdnet} focused training on few but higher quality labelled data, \gls{perchv1} opts for more data with potentially more noisy labels.
The model was trained to classify all levels of the taxonomy simultaneously (species, genus, family, and order).
As for \gls{birdnet}, embeddings for the \gls{perchv1} models are also obtained from the final pooling layer.

\subsubsubsection{Surfperch}
\Textcite{williams2024surfperch} continued the work of \textcite{google2023perch} by further pretraining the \gls{perchv1} model on marine bioacoustic data.
This model is referred to as \gls{surfperch}.
The embeddings for the \gls{surfperch} model are obtained as for \gls{perchv1}.

\subsubsubsection{Perch 2.0}
While \gls{perchv1} included only avian recordings, the \gls{perchv2} model incorporates a large multi-taxa dataset.
\Textcite{Merrienboer2025perchv2} trained the model using the even larger \gls{efficientnet-b3} \ac{cnn} architecture on a combination of four labelled audio datasets: \gls{xenocanto}, \gls{inaturalist}~\autocite{vanhorn2018inaturalist}, the \gls{tierstimmenarchiv}~\autocite{frommolt2006tierstimmenarchiv}, and \gls{fsd50k}~\autocite{Fonseca2022FSD50K}.
The model employs a novel training methodology that combines supervised classification with self-distillation using a prototype-learning classifier and an auxiliary source-prediction training criterion.
The authors show that embeddings produced by \gls{perchv2} exhibit strong linear separability in downstream tasks when compared to several other bioacoustic-specific embedding models, making \gls{perchv2} particularly effective for few-shot learning, clustering, and retrieval tasks across diverse bioacoustic domains.
Notably, despite containing almost no marine training data, \gls{perchv2} outperforms specialised marine models on marine transfer learning tasks~\autocite{Merrienboer2025perchv2}.
Unlike the other bioacoustic models considered in this work, it is uncertain whether \gls{perchv2} is strictly out-of-species with respect to elephant vocalisations.
Both the \gls{tierstimmenarchiv} and \gls{inaturalist} contain recordings of \textit{Loxodonta africana} and \textit{Elephas maximus}, so a small number of elephant recordings may be present in the \gls{perchv2} training corpus.
Because neither the full list of training samples nor the per-species taxonomy used for pretraining has been publicly disclosed, the precise extent of any overlap cannot be determined.
\Gls{perchv2} is therefore the only embedding model evaluated in this work whose status as out-of-domain or out-of-species cannot be established with certainty.
We have, however, manually verified that none of the recordings in the two evaluation datasets~(\cref{sec:data}) appear in the publicly accessible \gls{tierstimmenarchiv} or \gls{inaturalist} source corpora from which \gls{perchv2}'s training samples could have been drawn.

\subsubsubsection{AVES}
\Textcite{hagiwara2022aves} introduces four \gls{hubert}-based models: \gls{aves-core}, \gls{aves-bio}, \gls{aves-nonbio}, and \gls{aves-all}.
Each is trained using self-supervision~\autocite{hsu2021hubert} on a different combination of the \gls{fsd50k}, \gls{audioset}, and \gls{vggsound}~\autocite{Chen2020Vggsound} datasets.
The \gls{aves-bio} model is trained only on portions of \gls{audioset} annotated as containing animal sounds, along with \gls{fsd50k} and \gls{audioset} core sets.
Embeddings are obtained from the final transformer layer in the \gls{hubert} architecture.

\subsubsubsection{BirdAVES}
The Earthspecies project~\autocite{hagiwara2024birdaves} has most recently released three additional models: \gls{birdaves-biox-base}, \gls{birdaves-biox-large}, and \gls{birdaves-bioxn-large}.
These models have been trained on the same data as \gls{aves-bio}, with the addition of the \gls{xenocanto} and \gls{inaturalist} datasets.
Furthermore, the Earthspecies project has increased the model size from the \gls{hubert}[-base] to the \gls{hubert}[-large] configuration.
Embeddings are obtained as for \gls{aves}.

\subsubsubsection{animal2vec}
\Textcite{Zimmermann2026Animal2vecMeerKAT} introduce \gls{animal2vec}, a transformer applied directly to the raw audio waveform rather than to a spectral representation.
The waveform frontend is a learnable parametric \gls{sincnet} filterbank~\autocite{ravanelli2018sincnet}, followed by a stack of 1D convolutions.
Pretraining follows the data2vec~2.0 self-distillation scheme~\autocite{baevski2023data2vec2}: a student transformer, shown a masked view of the input, is trained to regress at the masked positions the contextualised representations produced on the unmasked input by a teacher transformer whose weights are an exponential moving average of the student's.
Pretraining is performed on the \gls{meerkat} dataset, consisting of \qty{1068}{hours} of meerkat~(\textit{Suricata suricatta}) vocalisations sampled at \qty{8}{\kilo\hertz}, of which \qty{184}{\hour} are strongly labelled at millisecond resolution across eleven classes.
Embeddings are obtained from the final transformer layer.

\subsection{Classification experiments}
\label{sec:shallow-classifier}
We perform multi-class single-label elephant call classification at the segment level.
Each endpointed call segment~(\cref{sec:segmentation}) is assigned to exactly one vocalisation class drawn from the call taxonomy of the respective dataset~(33~classes for \acs{elev}; 16 for \acs{ldc}; see \cref{tab:data:summary}).
Although the embedding models described in \cref{sec:emb-model} produce a sequence of per-frame embedding vectors, the classifier\footnotemark{} consumes the full sequence and emits a single classification per segment.
\footnotetext{Commonly referred to as a ``classification probe'' in the representation learning literature.}
Each classifier has a softmax output to produce the postulated vocalisation probability, with cross-entropy as the chosen loss function.

All classification experiments in this section were performed for each of the two datasets described in \cref{sec:data} using K-fold nested cross-validation~\autocite{mosteller1968crossval, stone1974crossval}.
In this procedure, each dataset is partitioned into $K$ disjoint subsets, referred to as \textit{folds}.
Each fold sequentially serves as a test set, whilst the remaining folds are used for training and development~(\textit{outer turn}).
For nested cross-validation, the development and training folds are similarly rotated~(\textit{inner turn}).
The final test results are computed by averaging the $K$ individual scores obtained for each test fold in the outer turn.

We consider \ac{lr} as a baseline.
\Ac{lr} is a linear classifier, and therefore provides a good indication of the degree to which the various embeddings separate the classes.
Next, we evaluate a \acf{mlp} classifier consisting of two fully connected layers, each with as many neurons as there are embedding dimensions, and a GELU activation function between the hidden layers.
This configuration was found to be effective in prior work on a comparable task~\autocite{geldenhuys2024aerd}.
For both \ac{lr} and the \ac{mlp}, when the embedding model produces a sequence of embedding vectors, we average these to produce a single vector of fixed dimensionality.
\Cref{fig:structure:flat} illustrates this classification procedure.

We also consider three recurrent architectures that are known to perform well at sequence tasks.
We evaluate the \gls{elman-rnn}~\autocite{elman1990rnn}, the \ac{gru}~\autocite{cho2014gru}, and the \ac{lstm}~\autocite{hochreiter1997lstm} as recurrent architectures.
\Cref{fig:structure:recurrent} illustrates the classification procedure followed for these recurrent neural architectures.
Each model is configured with one or two layers, a layerwise dropout between \qtyrange[range-phrase=\text{ and }]{0}{80}{\percent}, and either 32-, 64-, 128-, 256-dimensional hidden state.

Unless otherwise stated, each of the classification models is optimised using Adam~\autocite{kingma2014adam} with a peak learning rate ranging between \numrange[range-phrase=\text{ and }]{1e-3}{1e-6}, momentum terms of $\beta_1=0.9$ and $\beta_2=0.999$, and no weight decay.
Cross-entropy is used as the loss function for all experiments.

All combinations of the hyperparameters listed above were searched exhaustively for all embeddings, classifiers and datasets.
The final hyperparameter selection is determined by the lowest average developmental loss, computed over the inner development folds.

Finally, we employ early stopping by monitoring the training and development loss for either convergence or divergence.
If either loss has remained unchanged (convergence) or has repeatedly increased (divergence) for more than three epochs, training is halted.
The classification model weights are subsequently restored to those achieving the best development performance and evaluated against the test set.

\subsection{Performance evaluation}
We use the \acf{auc} as well as the \acf{map} as performance evaluation metrics.
These metrics are summative performance indicators over all classification thresholds, as opposed to measures calculated at a specific threshold.

The \ac{auc roc} curve is a widely used method for evaluating the performance of binary classifiers.
It is an aggregate measure across decision thresholds, and reflects the overall trade-off between false positives and false negatives offered by the classifier.
Since our task is multi-class, we extend the \ac{auc} by computing it for each call type individually and reporting the unweighted mean across all call types in the dataset.
Averaging in this way gives equal weight to each call type, so that rare call types contribute as much as common ones to the reported score; this is appropriate for our datasets, in which several call types are heavily under-represented.

The \ac{map} complements the \ac{auc} by summarising the trade-off between precision and recall.
The \ac{ap} for a particular call type is computed as the area under the precision-recall curve.
The \ac{map} is then obtained by averaging the per-call-type \ac{ap} scores with equal weight, in the same manner as for the \ac{auc}.

\section{Results}
In this section, we first consider elephant call classification using embeddings obtained from pretrained models.
Next, we analyse the classification performance achieved using features extracted from intermediate layers of the best embedding model.

\subsection{Classification using embedding models}
\label{sec:results:class}
In this section, we present the results of our elephant vocalisation classification experiments for different embedding and classification model pairs.
During training, only the classification model weights are optimised; the embedding model weights remain fixed.
Embeddings are therefore directly obtained from the pretrained models without any domain-specific adaptation.
A summary of the key results is presented in \cref{tab:results:class:summary}, whilst a complete set of results is presented in \cref{app:results:class}.

\begin{resultstable}
	\begin{table}[hp]
		\centering
		\caption{Summary of the key elephant vocalisation classification performance achieved using a subset of embedding models with the considered classifiers for the \acf{elev} and \acf{ldc} datasets.
			The \acf{auc} and \acf{map} are averaged over the outer~(test) cross-validation folds.
			The size of the embedding dimension~(\textit{Dim.}) along with the classification layer type~(\textit{Class.}) are shown.
			For each metric, the best-performing classifier within an embedding is underlined, whilst the best-performing embedding--classifier combination across all models is shown in bold.
		}
		\label{tab:results:class:summary}
\subheadinglabel{tab:results:class:benchmark}
		\subheadinglabel{tab:results:class:acoustic}
		\subheadinglabel{tab:results:class:asr}
		\subheadinglabel{tab:results:class:bio}
		\small
		\begin{tabular}{llllSSSSS}
			\toprule
			\resultsheading
\subheading{(\protect\subref{tab:results:class:benchmark}) Baseline}
\embedding[2]{\Acs{mfcc}}{40}                                                & GRU   &         & 0.7143     & \U{0.2177}   &              & \U{0.8091}   & \U{0.2489}                \\
			                                                                             &       & LSTM    &            & \U{0.7241}   & 0.1913       &              & 0.7728     & 0.2252       \\
			\newembed
\embedding[2]{\gls{beans40}~\autocite{hagiwara2022beans}}{160}               & LR    &         & 0.6786     & \U{0.2073}   &              & \U{0.8048}   & \U{0.2840}                \\
			                                                                             &       & MLP     &            & \U{0.6988}   & 0.1891       &              & 0.7848     & 0.2396       \\
			\newembed
\gls{aerd}~\autocite{geldenhuys2024aerd}                                     & N/A   & AST-seq &            & 0.8710       & {---}        &              & 0.9570     & {---}        \\
\subheading{(\protect\subref{tab:results:class:acoustic}) General audio}
\embedding[3]{\gls{vggish}~\autocite{hershey2017cnnaudio}}{128}              & Elman &         & \U{0.8286} & 0.2701       &              & 0.8735       & 0.3201                    \\
			                                                                             &       & GRU     &            & 0.8168       & 0.2849       &              & \U{0.8752} & \U{0.3215}   \\
			                                                                             &       & LSTM    &            & 0.8161       & \U{0.2965}   &              & 0.8734     & 0.3206       \\
			\newembed
\embedding[2]{\gls{beats}~(time+spec)~\autocite{chen2022beats}}{768}         & MLP   &         & 0.7879     & 0.2505       &              & \U{0.9057}   & 0.3202                    \\
			                                                                             &       & GRU     &            & \U{0.8137}   & \U{0.2816}   &              & 0.8991     & \U{0.4063}   \\
\subheading{(\protect\subref{tab:results:class:asr}) Speech}
\embedding[2]{\gls{w2v2-base}~\autocite{baevski2020wav2vec2}}{768}           & MLP   &         & \U{0.8306} & \U{0.2974}   &              & 0.8701       & 0.2763                    \\
			                                                                             &       & GRU     &            & 0.7803       & 0.2574       &              & \U{0.8773} & \U{0.3357}   \\
			\newembed
\embedding[2]{\gls{xlsr}~\autocite{conneua2020xlsr}}{1024}                   & LR    &         & 0.8009     & \U{0.2685}   &              & \U{0.8920}   & \U{0.2990}                \\
			                                                                             &       & MLP     &            & \U{0.8180}   & 0.2528       &              & 0.8826     & 0.2750       \\
			\newembed
\embedding[3]{\gls{hubert-base}~\autocite{hsu2021hubert}}{768}               & LR    &         & 0.8037     & 0.1982       &              & \U{0.8931}   & 0.2793                    \\
			                                                                             &       & MLP     &            & \U{0.8296}   & 0.2338       &              & 0.8885     & 0.2661       \\
			                                                                             &       & GRU     &            & 0.8221       & \U{0.2642}   &              & 0.8766     & \U{0.3221}   \\
\subheading{(\protect\subref{tab:results:class:bio}) Bioacoustics}
\embedding[3]{\gls{birdnetv2.4}~\autocite{kahl2021birdnet}}{1024}            & Elman &         & \U{0.8345} & 0.3635       &              & 0.9011       & 0.4064                    \\
			                                                                             &       & GRU     &            & 0.8095       & 0.3623       &              & \U{0.9082} & \U{0.4236}   \\
			                                                                             &       & LSTM    &            & 0.8172       & \U{0.3701}   &              & 0.9045     & 0.4090       \\
			\newembed
\embedding[4]{\gls{perchv1}~\autocite{google2023perch}}{1280}                & LR    &         & 0.8024     & 0.3628       &              & \U{0.9197}   & 0.4384                    \\
			                                                                             &       & Elman   &            & \U{0.8442}   & 0.3586       &              & 0.9155     & 0.4381       \\
			                                                                             &       & GRU     &            & 0.8433       & \U{0.3843}   &              & 0.9154     & 0.4400       \\
			                                                                             &       & LSTM    &            & 0.8356       & 0.3514       &              & 0.9103     & \U{0.4411}   \\
			\newembed
\embedding[3]{\gls{surfperch}~\autocite{williams2024surfperch}}{1280}        & LR    &         & 0.8189     & \U{0.4055}   &              & 0.9087       & \U{0.4516}                \\
			                                                                             &       & MLP     &            & 0.7309       & 0.2545       &              & \U{0.9160} & 0.4409       \\
			                                                                             &       & Elman   &            & \U{0.8434}   & 0.3795       &              & 0.9117     & 0.4385       \\
			\newembed
\embedding[3]{\gls{perchv2}~\autocite{Merrienboer2025perchv2}}{1536}         & MLP   &         & 0.7976     & 0.2929       &              & \B\U{0.9355} & 0.4757                    \\
			                                                                             &       & Elman   &            & \B\U{0.8492} & \B\U{0.4220} &              & 0.9314     & 0.5077       \\
			                                                                             &       & GRU     &            & 0.8453       & 0.3966       &              & 0.9312     & \B\U{0.5126} \\
			\newembed
\embedding[3]{\gls{birdaves-biox-base}~\autocite{hagiwara2024birdaves}}{768} & LR    &         & 0.8144     & 0.2284       &              & \U{0.8874}   & 0.2828                    \\
			                                                                             &       & MLP     &            & \U{0.8356}   & \U{0.2613}   &              & 0.8759     & 0.2772       \\
			                                                                             &       & GRU     &            & 0.7896       & 0.2336       &              & 0.8681     & \U{0.3411}   \\
\bottomrule
		\end{tabular}
	\end{table}
\end{resultstable}

First, we evaluate the baseline performance achieved by the classifiers using spectral features as embeddings~(\cref{sec:emb:baseline}).
Along with these baseline embeddings, a reference end-to-end supervised classification model~(\gls{aerd}) is provided~\autocite{geldenhuys2024aerd}.
\Cref{tab:results:class:benchmark} provides these baseline classification results.
When paired with the non-recurrent classifiers~(\ac{lr} and \ac{mlp}), the more elaborate \gls{beans} feature aggregation method yields performance comparable to simple \ac{mfcc} averaging on the \ac{elev} dataset, but substantially improves classification performance on the \ac{ldc} dataset.
However, the \acp{mfcc} become competitive when paired with a recurrent classifier.
Recurrent classifiers were not evaluated with \gls{beans} embeddings, as the inherent feature aggregation step removes temporal information.

Next, we perform a corresponding set of experiments using the embeddings obtained from general audio models, described in \cref{sec:emb:general}, and present the results in \cref{tab:results:class:acoustic}.
We observe that, even when a linear classifier is used, these embeddings show substantial performance improvements over the spectral features~(\cref{tab:results:class:benchmark}).
Within this category, \gls{beats} and \gls{vggish} achieve comparable performance on the \ac{elev} dataset, reaching \acp{auc} of \num{0.8137} and \num{0.8286}, respectively.
On the \ac{ldc} dataset, \gls{beats} achieves a clear advantage, with an \ac{mlp} classifier reaching an \ac{auc} of \num{0.9057} compared to \num{0.8752} for the best \gls{vggish} embedding-classifier combination.
Neither model, however, reaches the supervised \gls{aerd} baseline~(\acp{auc} of \num{0.8710} and \num{0.9570} on the \ac{elev} and \ac{ldc} datasets, respectively).

\Cref{tab:results:class:asr} shows the classification performance achieved when using transformer-based speech embedding models.
On the \ac{elev} dataset, the three speech models perform comparably, with \acp{auc} of \num{0.8306}, \num{0.8296}, and \num{0.8180} for \gls{wav2vec2}, \gls{hubert-base}, and \gls{xlsr}, respectively.
On the \ac{ldc} dataset, \gls{hubert-base} and \gls{xlsr} both reach an \ac{auc} close to \num{0.89}~(\num{0.8931} and \num{0.8920}, respectively), outperforming \gls{wav2vec2}~(\num{0.8773}) by approximately \qty{1.5}{\percent}~(absolute).
All three models share the same underlying transformer architecture but differ in their self-supervised training objectives and pretraining data.
Whilst the embeddings obtained from speech recognition models show improved classification performance compared to the baseline results~(\cref{tab:results:class:benchmark}), their performance is comparable to that achieved using general audio embedding models~(\cref{tab:results:class:acoustic}).

Finally, we evaluate the classification performance of embeddings obtained from bioacoustic models~(\cref{sec:emb:bio}), and present the results in \Cref{tab:results:class:bio}.
We note that, whilst these embedding models have been trained on bioacoustic data, this does not include the data described in \cref{sec:data}.
The bioacoustic models in this subsection are strictly out-of-species with respect to elephant vocalisations, with the exception of \gls{perchv2}.
As was described in \cref{sec:emb:bio}, the \gls{perchv2} training corpus may include elephant recordings drawn from the \gls{tierstimmenarchiv} or \gls{inaturalist}.
However, even if such recordings are present, they do not overlap with the two evaluation datasets used in this work~(\cref{sec:data}).
From the experimental results, embeddings exclusively trained on whalesong~(\textit{\gls{multi-whale}} and \textit{\gls{humpback}} models) lead to a performance degradation relative to the \mbox{\gls{mfcc}+\ac{lr}} baseline.
When the same underlying architecture~(\gls{efficientnet-b1}) is instead trained on terrestrial birdsong~(\textit{\gls{perchv1}}), or first on birdsong and then further on whalesong~(\textit{\gls{surfperch}}), we observe much better classification performance.

On the \ac{elev} dataset, \gls{perchv2} paired with an Elman classifier achieves the highest overall \ac{auc} across all embeddings~(\num{0.8492}), approaching but not exceeding the end-to-end supervised \gls{aerd} baseline~(\num{0.8710}).
\Gls{birdnet}, \gls{perchv1}, and \gls{perchv2} likewise show strong performance on the \ac{ldc} dataset, achieving \acp{auc} of \num{0.9082}, \num{0.9197}, and \num{0.9355}, respectively.
\Gls{perchv2} again achieves the best performance among all embedding models considered, approaching but not exceeding the \gls{aerd} baseline~(\num{0.9570}).
If \gls{perchv2} is set aside, since it is unclear whether its training set included elephant vocalisations, \gls{perchv1} becomes the top-performing strictly out-of-species model on both datasets, achieving \acp{auc} of \num{0.8442}~(Elman classifier) on the \ac{elev} dataset and \num{0.9197}~(\ac{lr} classifier) on the \ac{ldc} dataset.
This is within approximately \qty{2.7}{\percent} and \qty{3.7}{\percent}~(absolute) of the \gls{aerd} baseline, respectively.
The Perch family of models are therefore among the best-performing overall and superior to several transformer-based alternatives.
\Gls{surfperch}, which extends \gls{perchv1} with additional marine bioacoustic pretraining, also demonstrates strong performance, achieving an \ac{auc} of \num{0.9160} on the \ac{ldc} dataset.

Larger configurations of both the speech and bioacoustic model families were also evaluated~(\cref{app:results:class}).
It was found that the effect of increased model capacity on transfer performance differs markedly between the two domains.
For the speech models, performance deteriorates relative to their base counterparts.
For the \ac{elev} dataset, the \ac{auc} for \gls{wav2vec2} falls from \num{0.8773}~(base) to \num{0.8121}~(large), and for \gls{hubert} it falls from \num{0.8931}~(base) to \num{0.8566}~(large) and \num{0.8607}~(extra-large).
A similar degradation is observed for the \ac{ldc} dataset.
By contrast, scaling the BirdAVES family from \gls{birdaves-biox-base} to \gls{birdaves-biox-large} yields a modest improvement, with the best \ac{elev} \ac{auc} rising from \num{0.8356} to \num{0.8437} and the best \ac{ldc} \ac{auc} from \num{0.8874} to \num{0.8913}.

When comparing model rankings across evaluation metrics, we observe that the advantage of bioacoustic embeddings over their general audio and speech counterparts is more apparent in terms of \ac{map} than \ac{auc}.
On the \ac{elev} dataset, \gls{perchv2} leads both metrics, with an \ac{auc} of \num{0.8492} and an \ac{map} of \num{0.4220}, compared to \num{0.8306} and \num{0.2974} for the best non-bioacoustic alternative~(\gls{w2v2-base}).
The difference in \ac{auc} of approximately \qty{1.9}{\percent}~(absolute) is accompanied by a corresponding difference in \ac{map} of more than \qty{12.4}{\percent}~(absolute).
A similar pattern is observed on the \ac{ldc} dataset, where \gls{perchv2} again achieves the highest \ac{auc}~(\num{0.9355}), exceeding the best non-bioacoustic alternative~(\gls{beats} at \num{0.9057}) by approximately \qty{3.0}{\percent}~(absolute).
The same \gls{perchv2} model achieves an \ac{map} of \num{0.5126}, exceeding the best non-bioacoustic \ac{map}~(\num{0.4063} from \gls{beats}) by more than \qty{10.6}{\percent}~(absolute).

\begin{figure}[thp]
	\centering
	\begin{subfigure}{1.0\linewidth}
		\centering
		\caption{
			\Acs{elev} dataset
		}
		\label{fig:roc:elev}
		\includegraphics[width=0.9\textwidth]{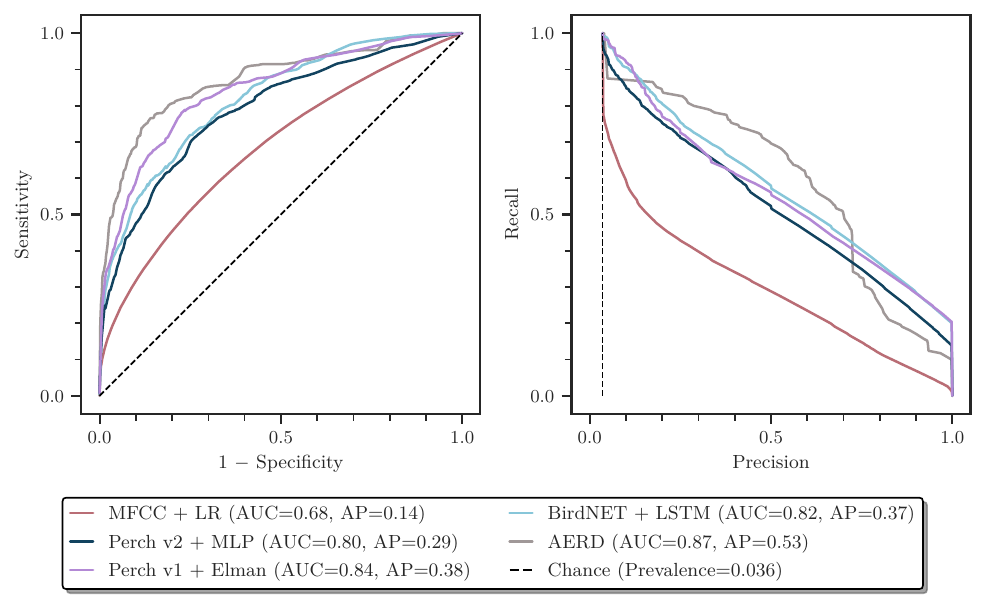}
	\end{subfigure}
	\begin{subfigure}{1.0\linewidth}
		\centering
		\caption{
			\Acs{ldc} dataset
		}
		\label{fig:roc:ldc}
		\includegraphics[width=0.9\textwidth]{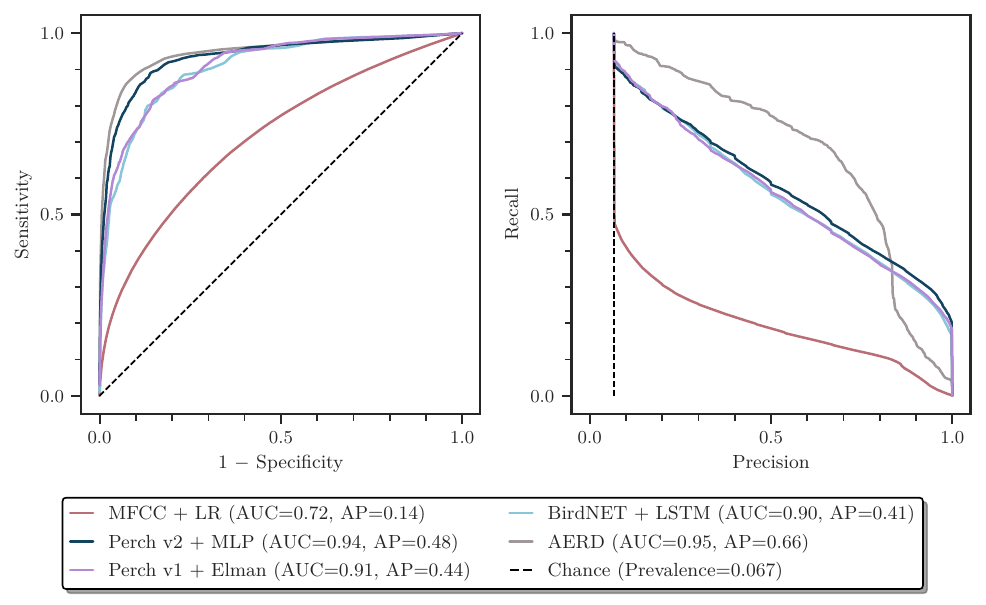}
	\end{subfigure}
	\caption{
		\Acf{roc} curves~(left) and precision-recall curves~(right) for a selection of embedding-classifier combinations evaluated on the \acs{elev} and \acs{ldc} datasets.
		The selection comprises the top-performing bioacoustic embedding-classifier pairs identified in \cref{tab:results:class:summary}, together with the \gls{mfcc}-based \acf{lr} baseline and the end-to-end supervised \gls{aerd} reference, both of which are included for comparison.
		For each combination, the \acf{auc} and \acf{ap} values reported in the legend are averaged across the outer~(test) cross-validation folds.
		The diagonal dashed line~(\textit{Chance}) shown in the \ac{roc} panels marks the expected performance of a random classifier.
		The horizontal dashed line~(\textit{Prevalence}) shown in the precision-recall panels marks the precision achieved by a classifier deciding at random according to the class prior.
	}
	\label{fig:roc}
\end{figure}

These differences between \ac{auc} and \ac{map} are also reflected in the \acf{roc} and precision-recall curves shown in \cref{fig:roc}.
The \ac{roc} curves for the bioacoustic embedding-classifier combinations track the \gls{aerd} reference closely, with \gls{perchv2} reaching an \ac{auc} of \num{0.94} on the \ac{ldc} dataset, compared to \num{0.95} for \gls{aerd}.
The corresponding precision-recall curves, however, reveal a clearer separation, with \gls{aerd} retaining a clear advantage in \ac{ap} on both datasets.
This advantage is most pronounced on the \ac{ldc} dataset, where the \gls{aerd} curve lies above those of the embedding-based models across most of the recall range, exceeding the best embedding by approximately \num{0.18} in \ac{ap}.

\subsection{Layerwise analysis}

In this section, we present the elephant vocalisation classification results achieved when embeddings are obtained from intermediate transformer layer representations, instead of from the final layer as done in the previous section.
We choose to limit our experiments to the top-performing transformer models identified in \cref{sec:results:class}.
Although these are not the strongest performers overall~(since the leading bioacoustic embeddings all use a \ac{cnn} architecture), a layerwise analysis is meaningful only for transformer architectures.
We therefore restrict the comparison to transformers of comparable~(base) parameter count.
Additionally, we also evaluate the representations that constitute the input to the first transformer layer, referred to as \textit{input features}.
In the case of \gls{wav2vec2} and \gls{hubert}, these features are obtained from the final \ac{cnn} layer in the feature extractor stage of the model architecture.
For \gls{beats}, these features take the form of a linear projection of the log-scale mel spectrogram.
Furthermore, the chosen models are limited to base-sized variants consisting of twelve transformer encoder layers, each of which includes the self-attention mechanism.
We only evaluate \ac{lr} as the classification model in these experiments, as linear probing is the established methodology for layerwise analysis~\autocite{pasad2021layerwise}.
The objective is to assess the linear separability of each layer's representations, rather than to maximise classification performance.

\begin{figure}
	\centering
	\begin{subfigure}{1.0\linewidth}
		\centering
		\caption{
			\Acs{elev} dataset
		}
		\label{fig:layerwise:elev}
		\includegraphics{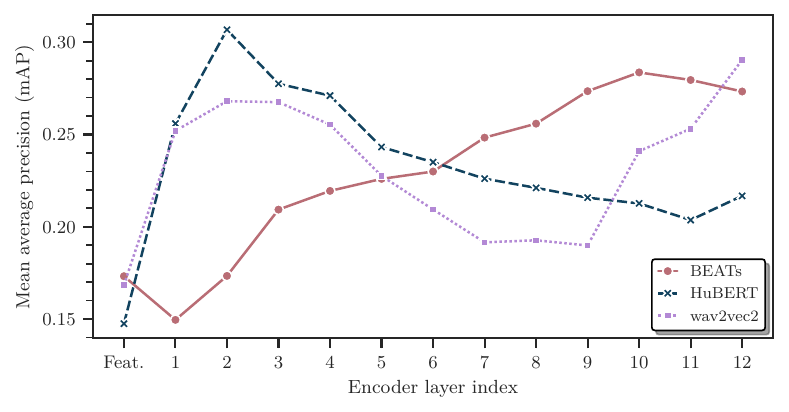}
	\end{subfigure}
	\begin{subfigure}{1.0\linewidth}
		\centering
		\caption{
			\Acs{ldc} dataset
		}
		\label{fig:layerwise:ldc}
		\includegraphics{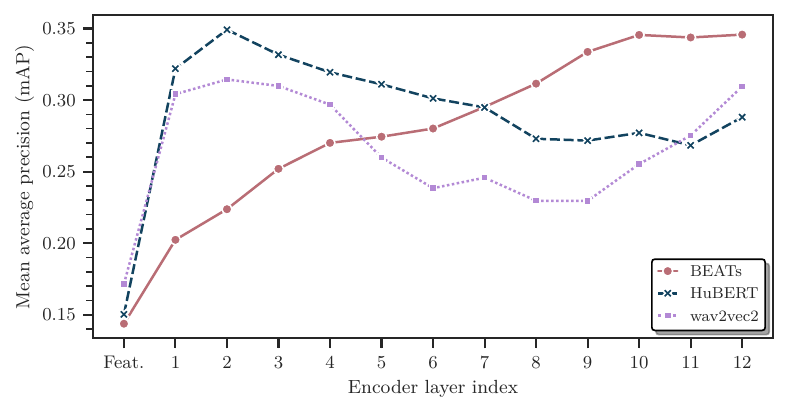}
	\end{subfigure}
	\caption{
		Layerwise classification performance given by the \acf{map} computed for elephant vocalisations drawn from the development folds, for a single outer turn.
		Embeddings are obtained from intermediate transformer layers, as well as input features~(\textit{Feat.}) to the initial transformer layer.
		In the case of \gls{hubert} and \gls{wav2vec2} the input transformer features are obtained from the final \ac{cnn} layer, whereas for the \gls{beats} model this is the log-mel spectrogram.
		The selected embedding models are all the \textit{base} configuration size, with only \acf{lr} considered as the classification model.
	}
	\label{fig:layerwise}
\end{figure}

\Cref{fig:layerwise} shows the results for these layerwise experiments.
We observe similar characteristic behaviour to the layerwise analysis conducted by \textcite{hsu2021hubert, chen2022wavlm} for speech-related tasks.
This is seen in the characteristic \emph{U-shape} in performance, where earlier encoder layers show strong classification performance, followed by a performance decrease at the central layers and an increase in performance at the final layers~(in the case of \gls{wav2vec2}).
Although this trend is observed for \gls{wav2vec2} over both datasets, it was not seen for \gls{beats}, where deeper layers lead to almost linearly improved classification performance.
Beyond a certain depth, however, no further improvement is observed.

Although the second transformer layer of \gls{hubert} yielded the strongest performance within that architecture on both datasets, and \gls{wav2vec2} reached its peak at either the second or final transformer layer depending on the dataset, \gls{beats} exhibited no comparable layerwise peak.
Across all three models, however, no intermediate layer surpasses the leading final-layer bioacoustic embeddings~(\gls{perchv2} and \gls{perchv1}) reported in \cref{sec:results:class}.

Lastly, we observe that the \ac{cnn} features of the \gls{wav2vec2} and \gls{hubert} models perform roughly as well as the mel~spectrogram features produced by the \gls{beats} model, with the latter performing slightly better on the \gls{elev} dataset.

\section{Discussion}
Our central hypothesis, that pretrained embedding models encode acoustic information that generalises to elephant call classification without domain-specific fine-tuning, is supported by our experimental results.
Across both datasets, embeddings obtained from pretrained speech, general audio, and bioacoustic models achieve classification performance that in some cases approaches the end-to-end supervised baseline, despite no fine-tuning of the embedding models being performed.
In all cases considered, the embedding models offer improvements over the spectral feature baselines~(\gls{mfcc} and \gls{beans}).

We find that in most scenarios, averaging the embeddings resulting from transformer-based models, and applying a linear classification model achieves strong performance.
Recurrent classifiers were found to offer further improvements, especially when paired with \ac{cnn}[-based] embeddings.
We believe this is because the transformer-based embedding models have already modelled the long- and short-range dependencies using the self-attention mechanism.
In contrast, the receptive fields of the \ac{cnn} embedding models cannot capture long-range dependencies well, and thus benefit from the addition of a sequence-based model, such as a recurrent network.

The three speech models, although closely related in architecture, diverged in transfer performance.
\Gls{xlsr} retained a clear advantage over \gls{wav2vec2} on the \ac{ldc} dataset, consistent with its pretraining on a larger and more diverse multilingual speech corpus that exposes the model to a wider range of speakers and acoustic environments than the LibriSpeech corpus used by \gls{wav2vec2}.
The LibriSpeech corpus used to pretrain \gls{wav2vec2} and \gls{hubert}, by contrast, consists of high-quality recordings made in controlled or near-studio conditions, which do not reflect the field-recording environment from which elephant vocalisations are gathered.
\Gls{hubert-base} likewise outperformed \gls{wav2vec2} on the \ac{ldc} dataset despite both being pretrained on LibriSpeech, suggesting that the quantised codebook objective used to train \gls{wav2vec2} may yield representations that are less informative for non-speech signals.

The strongest embedding-classifier combinations approach, but do not exceed, the end-to-end supervised \gls{aerd} baseline in terms of \ac{auc}.
On the \ac{elev} dataset, \gls{perchv2} with an Elman classifier achieves an \ac{auc} of \num{0.8492}, within \qty{2.2}{\percent} of the \gls{aerd} baseline of \num{0.8710}.
On the \ac{ldc} dataset, \gls{perchv2} paired with an \gls{mlp} classifier achieves an \ac{auc} of \num{0.9355}, again within \qty{2.2}{\percent} of the \gls{aerd} baseline of \num{0.9570}.
This result is notable given that \gls{aerd} is explicitly trained for elephant call detection and classification, whereas \gls{perchv2} was pretrained on a predominantly non-elephant corpus.
Because the precise composition of its training set is unknown and might include a small number of elephant calls~(\cref{sec:emb:bio}), \gls{perchv2} cannot with complete certainty be considered strictly out-of-species.
Setting \gls{perchv2} aside on this basis, \gls{perchv1}, which was trained only on avian vocalisations, becomes the strongest strictly out-of-species model, reaching \acp{auc} of \num{0.8442}~(Elman) on the \ac{elev} dataset and \num{0.9197}~(\ac{lr}) on the \ac{ldc} dataset, within approximately \qty{2.7}{\percent} and \qty{3.7}{\percent}~(absolute) of the \gls{aerd} baseline, respectively.

Among the bioacoustic models, the strongest transfer was achieved by those pretrained on terrestrial vocalisations.
\Gls{perchv1} and \gls{birdnet}, both trained on avian song, generalised effectively to elephant calls on both datasets, whereas the marine-only models~(\gls{humpback} and \gls{multi-whale}) performed below the spectral baselines.
We attribute this contrast to the substantial differences between marine and terrestrial soundscapes in propagation medium, ambient noise, and the spectral properties of the recorded vocalisations.
\Gls{surfperch}, however, which extends \gls{perchv1} with additional marine pretraining, retained strong performance on both elephant datasets.
This indicates that marine acoustic data can be a valuable source of pretraining diversity once a foundation of terrestrial representations has been established, even though marine pretraining alone does not transfer to a terrestrial target.

Increasing model capacity did not yield consistent gains across the embedding families.
The larger variants of \gls{wav2vec2} and \gls{hubert} performed worse than their base counterparts on both datasets, in agreement with \textcite{Cauzinille2025SpeechTransferBio}, who report a similar reduction in transferability for higher-capacity speech models when the downstream task is data-scarce.
By contrast, scaling the BirdAVES family from base to large yielded modest gains on both the \ac{elev} and \ac{ldc} datasets.
We interpret this contrast as indicating that the benefit of additional capacity is contingent on alignment between the pretraining and downstream domains: gains emerged only when the pretraining data were acoustically relevant to the target task.

When evaluated using \ac{map}, the bioacoustic embedding models consistently achieve the strongest results.
On the \ac{ldc} dataset, all of the five top-ranked embedding-classifier combinations by \ac{map} correspond to \gls{perchv2}.
Conversely, speech and general audio models, though competitive in \ac{auc}, are weaker when measured by \ac{map}.
\Gls{beats} represents the clearest example: with a \gls{gru} classifier, it achieves an \ac{auc} of \num{0.8991} on the \ac{ldc} dataset~(within approximately \qty{4}{\percent} of the best-performing embedding), yet its \ac{map} of \num{0.4063} is more than \qty{10}{\percent} below that of the leading bioacoustic embeddings.
This gap reflects the greater sensitivity of \ac{map} to false positives on rare call types.
The stronger \ac{map} results achieved by the bioacoustic models therefore suggest that these models offer more balanced classification performance across all call types, not only those that are well represented.
\Cref{fig:roc} confirms this pattern, with the gap between the embedding-based models and the \gls{aerd} reference more pronounced on the precision-recall curves than on the \ac{roc} curves.

At low recall, the leading embedding-based models attain higher precision than the \gls{aerd} reference on both datasets~(see \cref{fig:roc}).
This advantage diminishes as higher recall is required, with \gls{aerd} retaining its overall lead in \ac{ap}.
In deployment settings where high precision is required and some missed calls are acceptable, a pretrained-embedding classifier may therefore be preferable to an end-to-end fine-tuned model.
Several scenarios common in elephant bioacoustics fit this profile.
The mining of large \ac{pam} archives is one such case, since exhaustive recall is unattainable through manual review and the value of automation lies in high-confidence detections for expert verification.
The same logic extends to occupancy and presence-absence surveys, which underpin much of biodiversity monitoring and require only a single reliable detection to confirm that a species has visited an area.
Studies of individual identification from rumble structure~\autocite{Pardo2024ElephantNameAddress} and of calls associated with specific behaviours~\autocite{poole1988social,Soltis2005ElephantVocalShortRange,Poole2005VocalLearning} also depend on reliable exemplars rather than on exhaustive coverage.
This balance is most consequential in human-elephant conflict early-warning systems, in which alerts are issued to villagers and rangers when approaching elephants are detected, and false alarms readily produce alarm fatigue and the loss of community trust.

Our layerwise analysis of the transformer embedding models revealed strong agreement with similar experiments in human speech processing~\autocite{pasad2021layerwise}.
The early transformer layers of \gls{wav2vec2} and \gls{hubert} are understood to encode low-level acoustic features, with deeper layers progressively capturing phoneme-, word-, and intent-level structure.
Although elephant vocalisations and human speech are very different communication systems, the speech-pretrained models displayed the same characteristic U-shape under linear probing, with peak \ac{map} recovered at intermediate or final layers rather than at the deepest layer.
Within this pattern, \gls{hubert} attains its peak at the second transformer layer, whereas \gls{wav2vec2} only reaches its peak at the final layer, reinforcing the suggestion that the quantised codebook objective yields less directly usable intermediate representations.
\Gls{beats}, by contrast, was pretrained on \gls{audioset} without a linguistic objective, and its linear-probe performance increased monotonically with depth.
We attribute this to the absence of a hierarchical linguistic structure in the \gls{beats} pretraining task: general acoustic separability refines with depth, without acquiring the linguistic hierarchy seen in the speech models.
Across all three models, however, no intermediate layer surpassed the leading final-layer bioacoustic embeddings reported in \cref{sec:results:class}.
Unfortunately, due to the very limited extent of the annotations, it was not possible to further investigate whether a call structure hierarchy similar to that found in human speech exists also for elephants.
This underscores the need for open datasets that contain vocal unit or note-level annotations for elephant calls~\autocite{poole1988social,Soltis2005ElephantVocalShortRange,Kershenbaum2016VocalUnits}.
Such a dataset would allow further analysis of the similarities between the structures of elephant calls and human speech.

\section{Conclusion}
This study presents the first comprehensive evaluation of the use of audio embedding models for elephant vocalisation classification, where the embedding models have been pretrained on data that is either out-of-domain or out-of-species.
This addresses the often-encountered constraint in computational bioacoustics where traditional machine learning approaches are constrained by the scarcity and high annotation cost of expert-labelled data.
We hypothesised that the high-dimensional numerical vectors obtained from models pretrained on diverse acoustic domains~(including general audio, human speech, and bioacoustic recordings) would encode information that generalises to elephant communication, even without domain-specific fine-tuning.
Our extensive experiments across two elephant species datasets validate this hypothesis and reveal the following insights.
Classification experiments demonstrate that pretrained embeddings can achieve performance approaching that of end-to-end supervised models specifically trained for elephant call detection and classification, without any fine-tuning of the embedding model.
The \gls{perchv2} model achieved the strongest performance overall, with an \ac{auc} of \num{0.8492}~(using an Elman recurrent classifier) for African bush elephant~(\textit{Loxodonta africana}) call classification, and an \ac{auc} of \num{0.9355}~(using an \ac{mlp} classifier) for Asian elephant~(\textit{Elephas maximus}) call classification.
\Gls{perchv1}, the precursor to \gls{perchv2}, achieves almost identical performance~(\acp{auc} of \num{0.8442} and \num{0.9197} on the respective datasets) despite its training corpus containing no elephant vocalisations.
Both Perch models substantively outperform baseline spectral features and demonstrate that embedding-based approaches can serve as viable alternatives to end-to-end supervised training, even when pretrained on out-of-domain and out-of-species data.

A comparative analysis across embedding domains provides insights into acoustic feature transferability and domain adaptation.
General audio models, particularly \gls{beats} and \gls{vggish}, showed substantial performance improvements over spectral baselines.
Although no embedding-classifier combination exceeded the end-to-end supervised \gls{aerd} model in terms of \ac{auc}, the best-performing combinations came within \qty{2.2}{\percent} of this baseline.
The precision-recall curves show a wider separation, however, with the \gls{aerd} reference retaining a clear lead in \ac{ap}, particularly on the \ac{ldc} dataset.
This is notable given that the majority of the embedding models were not pretrained on elephant vocalisations, whereas \gls{aerd} is explicitly supervised on elephant-specific data.
These findings suggest that self-supervised pretraining on large-scale audio corpora captures acoustic representations that generalise effectively to specialised bioacoustic domains, even without any in-domain fine-tuning.

Speech-based transformer models performed competitively with general audio models, suggesting that self-supervised representations learned from human speech capture acoustic features that generalise to other vocalisation types, including elephant communication.
This cross-domain effectiveness is consistent with recent findings by other researchers showing transfer learning success from speech models to other animal vocalisations~\autocite{Cauzinille2025SpeechTransferBio}, though further research is needed to determine whether this reflects shared communicative structures or more general acoustic properties.
Interestingly, our results contrast with findings by \textcite{Cauzinille2025SpeechTransferBio}, who observed that linear classifiers performed comparably to recurrent architectures across diverse bioacoustic tasks.
In our experiments, recurrent classifiers consistently outperformed linear models, suggesting that the sequential structure of elephant vocalisations may benefit from temporal modelling that linear classifiers cannot capture.
Given that elephant societies do exhibit the ability for vocal learning, and as a result present structured communication mechanisms~\autocite{Poole2005VocalLearning,Stoeger2014VocalLearning}, we speculate that these self-supervised representations may be capturing relevant acoustic structures.

Bioacoustic models trained on avian data demonstrated exceptional and consistent performance across both \ac{auc} and \ac{map}, with \gls{perchv1} and \gls{birdnet} achieving \acp{auc} exceeding \num{0.90} on the \ac{ldc} dataset.
The recently released \gls{perchv2} model, which incorporates multi-taxa training data, achieved an even higher \ac{auc} of \num{0.9355} on the \ac{ldc} dataset.
The strong performance of the strictly out-of-species avian models, particularly \gls{perchv1}, validates our hypothesis regarding out-of-species generalisation within terrestrial domains, and suggests that acoustic environment similarity plays a crucial role in feature transferability.
Conversely, marine bioacoustic models (\gls{humpback}, \gls{multi-whale}) showed degraded performance relative to baseline features, emphasising that acoustic environment similarity remains critical.

A layerwise analysis, which constitutes the first investigation of intermediate transformer representations for elephant call classification, reveals patterns remarkably consistent with hierarchical feature learning observed in human speech processing tasks.
The characteristic performance curves across transformer layers~(with strong early layer performance, mid-layer degradation, and late-layer recovery) lead us to speculate that elephant vocalisations exhibit structured acoustic hierarchies analogous to human speech.
The second layer of both \gls{wav2vec2} and \gls{hubert} provided sufficient information for effective classification, indicating that approximately \qty{10}{\percent} of the parameters of the full pretrained model can be retained whilst maintaining strong performance, albeit without exceeding the best avian bioacoustic embeddings.
This might enable efficient edge deployment for real-time conservation applications.

Key limitations of our study include the lack of fine-grained call annotations, which prevents detailed analysis of communication hierarchies, and the absence of contextual information (environment, social context, temporal patterns) that could enhance both downstream performance and ecological relevance.

In conclusion, this work establishes pretrained embedding models as a powerful and practical approach to elephant vocalisation classification, offering both immediate conservation applications and fundamental insights into acoustic feature transferability across domains and species.
The demonstrated effectiveness of out-of-domain and out-of-species embeddings challenges traditional assumptions about the necessity of task-specific training data, opening new possibilities for data-efficient bioacoustic research.
As the field continues to develop larger and more sophisticated pretrained models, the approach validated in this study provides a foundation for advancing computational bioacoustics and supporting critical wildlife conservation efforts through automated monitoring and analysis systems.

\section*{Acknowledgements}
We gratefully acknowledge financial support by Telkom~(South Africa) for the research presented in this paper.
We would also like to thank the Stellenbosch Rhasatsha \ac{HPC} facility and its team for access to their facilities and technical support.
In particular, we honour the memory of the late Gerhard van Wageningen, former Head of Research Infrastructure and \ac{HPC}, whose dedication was instrumental in establishing and maintaining the computing resources on which this work relied.

\section*{ORCID}
\begin{description}
	\item[Christiaan M. Geldenhuys] \orcidlinkf{0000-0003-0691-0235}
	\item[Thomas R. Niesler] \orcidlinkf{0000-0002-7341-1017}
\end{description}

\clearpage
\newrefcontext[sorting=nyt] \printbibliography

\clearpage
\appendix
\begin{appendices}
	\section{Complete embedding classification results}
	\label{app:results:class}
	\begin{resultstable}
	\begin{longtable}[c]{llll*{5}{S}}
\caption{
			Elephant vocalisation classification performance achieved using embedding models~(\cref{sec:emb}) with the considered classifiers for the \acf{elev} and \acf{ldc} datasets.
			The \acf{auc} and \acf{map} are averaged over the outer~(test) cross-validation folds.
			The size of the embedding dimension~(\textit{Dim.}) along with the classification layer type~(\textit{Class.}) are shown.
			Hyperparameter optimisation has been performed using a grid search approach, for each embedding, classifier, and dataset combination.
			For each metric, the best-performing classifier within an embedding is underlined, whilst the best-performing embedding--classifier combination across all models is shown in bold.
		}
		\label{tab:results:class:appendix}
		\\ \toprule
		\resultsheading
		\endfirsthead
		\multicolumn{9}{r}{{\small \textit{\tablename~\thetable~--~continued}}}                                                                                                             \\
		\toprule
		\resultsheading
		\midrule
		\endhead
		\multicolumn{9}{r}{{\small \textit{Continued on next page}}}                                                                                                                        \\
		\endfoot
		\endlastfoot
\subheading{(a)~Baseline}
\embedding{\Acs{mfcc}}{40}                                                    & LR  &         & 0.6814     & 0.1418       &              & 0.7162     & 0.1359                      \\*
		                                                                              &     & MLP     &            & 0.6897       & 0.1504       &            & 0.7131       & 0.1408       \\*
		                                                                              &     & Elman   &            & 0.6475       & 0.1432       &            & 0.6040       & 0.0994       \\*
		                                                                              &     & GRU     &            & 0.7143       & \U{0.2177}   &            & \U{0.8091}   & \U{0.2489}   \\*
		                                                                              &     & LSTM    &            & \U{0.7241}   & 0.1913       &            & 0.7728       & 0.2252       \\*
		\newembed
\embedding[2]{\gls{beans20}~\autocite{hagiwara2022beans}}{80}                 & LR  &         & \U{0.7136} & \U{0.2230}   &              & \U{0.7947} & \U{0.2816}                  \\*
		                                                                              &     & MLP     &            & 0.7080       & 0.1899       &            & 0.7843       & 0.2731       \\*
		\newembed
\embedding[2]{\gls{beans40}~\autocite{hagiwara2022beans}}{160}                & LR  &         & 0.6786     & \U{0.2073}   &              & \U{0.8048} & \U{0.2840}                  \\*
		                                                                              &     & MLP     &            & \U{0.6988}   & 0.1891       &            & 0.7848       & 0.2396       \\*
		\newembed
\gls{aerd}~\autocite{geldenhuys2024aerd}                                      & N/A & AST-seq &            & 0.8710       & {---}        &            & 0.9570       & {---}        \\*
\subheading{(b)~General audio}
\embedding{\gls{vggish}~\autocite{hershey2017cnnaudio}}{128}                  & LR  &         & 0.7703     & 0.2151       &              & 0.7889     & 0.2154                      \\*
		                                                                              &     & MLP     &            & 0.7814       & 0.1832       &            & 0.7975       & 0.2150       \\*
		                                                                              &     & Elman   &            & \U{0.8286}   & 0.2701       &            & 0.8735       & 0.3201       \\*
		                                                                              &     & GRU     &            & 0.8168       & 0.2849       &            & \U{0.8752}   & \U{0.3215}   \\*
		                                                                              &     & LSTM    &            & 0.8161       & \U{0.2965}   &            & 0.8734       & 0.3206       \\*
		\newembed
\embedding{\gls{beats}~(time+spec)~\autocite{chen2022beats}}{768}             & LR  &         & 0.7951     & 0.2768       &              & 0.8991     & 0.3099                      \\*
		                                                                              &     & MLP     &            & 0.7879       & 0.2505       &            & \U{0.9057}   & 0.3202       \\*
		                                                                              &     & Elman   &            & 0.7955       & 0.2506       &            & 0.8760       & 0.3629       \\*
		                                                                              &     & GRU     &            & \U{0.8137}   & \U{0.2816}   &            & 0.8991       & \U{0.4063}   \\*
		                                                                              &     & LSTM    &            & 0.7235       & 0.1917       &            & 0.8590       & 0.3207       \\*
		\newembed
\embedding{\gls{beats}~(time)~\autocite{chen2022beats}}{768}                  & LR  &         & 0.7883     & 0.2688       &              & 0.9069     & 0.3419                      \\*
		                                                                              &     & MLP     &            & 0.8098       & \U{0.2843}   &            & \U{0.9092}   & 0.3397       \\*
		                                                                              &     & Elman   &            & 0.8073       & 0.2738       &            & 0.8870       & 0.3760       \\*
		                                                                              &     & GRU     &            & \U{0.8108}   & 0.2774       &            & 0.9030       & \U{0.4011}   \\*
		                                                                              &     & LSTM    &            & 0.7677       & 0.2569       &            & 0.8858       & 0.3788       \\*
		\newembed
\embedding[2]{\gls{beats}~(spec)~\autocite{chen2022beats}}{768}               & LR  &         & 0.7958     & 0.2729       &              & 0.8956     & \U{0.3946}                  \\*
		                                                                              &     & MLP     &            & \U{0.8178}   & 0.2630       &            & \U{0.9039}   & 0.3912       \\*
\subheading{(c)~Speech}
\embedding{\gls{w2v2-base}~\autocite{baevski2020wav2vec2}}{768}               & LR  &         & 0.8184     & 0.2846       &              & 0.8708     & 0.2946                      \\*
		                                                                              &     & MLP     &            & \U{0.8306}   & \U{0.2974}   &            & 0.8701       & 0.2763       \\*
		                                                                              &     & Elman   &            & 0.7889       & 0.2461       &            & 0.8602       & 0.2984       \\*
		                                                                              &     & GRU     &            & 0.7803       & 0.2574       &            & \U{0.8773}   & \U{0.3357}   \\*
		                                                                              &     & LSTM    &            & 0.7576       & 0.2241       &            & 0.8516       & 0.3025       \\*
		\newembed
\embedding{\gls{w2v2-large}~\autocite{baevski2020wav2vec2}}{1024}             & LR  &         & 0.7267     & 0.1693       &              & 0.8083     & 0.1984                      \\*
		                                                                              &     & MLP     &            & \U{0.7309}   & \U{0.1896}   &            & 0.7828       & 0.2013       \\*
		                                                                              &     & Elman   &            & 0.6440       & 0.1487       &            & 0.7383       & 0.1775       \\*
		                                                                              &     & GRU     &            & 0.7160       & 0.1511       &            & \U{0.8121}   & \U{0.2247}   \\*
		                                                                              &     & LSTM    &            & 0.6072       & 0.1083       &            & 0.7583       & 0.1808       \\*
		\newembed
\embedding{\gls{xlsr}~\autocite{conneua2020xlsr}}{1024}                       & LR  &         & 0.8009     & \U{0.2685}   &              & \U{0.8920} & \U{0.2990}                  \\*
		                                                                              &     & MLP     &            & \U{0.8180}   & 0.2528       &            & 0.8826       & 0.2750       \\*
		                                                                              &     & Elman   &            & 0.7623       & 0.1870       &            & 0.8400       & 0.2318       \\*
		                                                                              &     & GRU     &            & 0.7995       & 0.2055       &            & 0.8567       & 0.2465       \\*
		                                                                              &     & LSTM    &            & 0.7655       & 0.2257       &            & 0.8456       & 0.2481       \\*
		\newembed
\embedding{\gls{hubert-base}~\autocite{hsu2021hubert}}{768}                   & LR  &         & 0.8037     & 0.1982       &              & \U{0.8931} & 0.2793                      \\*
		                                                                              &     & MLP     &            & \U{0.8296}   & 0.2338       &            & 0.8885       & 0.2661       \\*
		                                                                              &     & Elman   &            & 0.8109       & 0.2347       &            & 0.8787       & 0.3162       \\*
		                                                                              &     & GRU     &            & 0.8221       & \U{0.2642}   &            & 0.8766       & \U{0.3221}   \\*
		                                                                              &     & LSTM    &            & 0.7946       & 0.2467       &            & 0.8636       & 0.3056       \\*
		\newembed
\embedding{\gls{hubert-large}~\autocite{hsu2021hubert}}{1024}                 & LR  &         & 0.7791     & \U{0.1951}   &              & 0.8560     & 0.2285                      \\*
		                                                                              &     & MLP     &            & \U{0.8032}   & 0.1911       &            & 0.8505       & 0.2263       \\*
		                                                                              &     & Elman   &            & 0.7182       & 0.1738       &            & 0.8416       & 0.2348       \\*
		                                                                              &     & GRU     &            & 0.7123       & 0.1568       &            & 0.8455       & 0.2474       \\*
		                                                                              &     & LSTM    &            & 0.7020       & 0.1560       &            & \U{0.8566}   & \U{0.2511}   \\*
		\newembed
\embedding{\gls{hubert-xlarge}~\autocite{hsu2021hubert}}{1280}                & LR  &         & 0.7813     & 0.1952       &              & 0.8390     & 0.2282                      \\*
		                                                                              &     & MLP     &            & \U{0.8133}   & \U{0.2006}   &            & \U{0.8607}   & \U{0.2413}   \\*
		                                                                              &     & Elman   &            & 0.6514       & 0.1388       &            & 0.8186       & 0.1883       \\*
		                                                                              &     & GRU     &            & 0.7470       & 0.1778       &            & 0.8268       & 0.2095       \\*
		                                                                              &     & LSTM    &            & 0.7224       & 0.1578       &            & 0.8399       & 0.2253       \\*
\subheading{(d)~Bioacoustics}
\embedding{\Gls{humpback}~\autocite{google2021humpback}}{2048}                & LR  &         & 0.5653     & 0.1276       &              & 0.8418     & \U{0.2868}                  \\*
		                                                                              &     & MLP     &            & 0.5543       & 0.1177       &            & 0.8451       & 0.2752       \\*
		                                                                              &     & Elman   &            & 0.7709       & 0.2112       &            & 0.8431       & 0.2440       \\*
		                                                                              &     & GRU     &            & 0.7854       & 0.2270       &            & \U{0.8621}   & 0.2711       \\*
		                                                                              &     & LSTM    &            & \U{0.7874}   & \U{0.2446}   &            & 0.8514       & 0.2468       \\*
		\newembed
\embedding{\Gls{multi-whale}~\autocite{google2024multiwhale}}{1280}           & LR  &         & 0.5622     & 0.1103       &              & 0.8531     & 0.2680                      \\*
		                                                                              &     & MLP     &            & 0.6686       & 0.1733       &            & 0.8664       & 0.2652       \\*
		                                                                              &     & Elman   &            & 0.7961       & 0.2614       &            & 0.8845       & 0.3244       \\*
		                                                                              &     & GRU     &            & \U{0.8275}   & \U{0.2851}   &            & \U{0.8940}   & \U{0.3410}   \\*
		                                                                              &     & LSTM    &            & 0.8094       & 0.2659       &            & 0.8835       & 0.3241       \\*
		\newembed
\embedding{\gls{birdnetv2.4}~\autocite{kahl2021birdnet}}{1024}                & LR  &         & 0.7507     & 0.3106       &              & 0.8788     & 0.3578                      \\*
		                                                                              &     & MLP     &            & 0.7834       & 0.3090       &            & 0.8894       & 0.3527       \\*
		                                                                              &     & Elman   &            & \U{0.8345}   & 0.3635       &            & 0.9011       & 0.4064       \\*
		                                                                              &     & GRU     &            & 0.8095       & 0.3623       &            & \U{0.9082}   & \U{0.4236}   \\*
		                                                                              &     & LSTM    &            & 0.8172       & \U{0.3701}   &            & 0.9045       & 0.4090       \\*
		\newembed
\embedding{\gls{perchv1}~\autocite{google2023perch}}{1280}                    & LR  &         & 0.8024     & 0.3628       &              & \U{0.9197} & 0.4384                      \\*
		                                                                              &     & MLP     &            & 0.7798       & 0.2983       &            & 0.9180       & 0.4385       \\*
		                                                                              &     & Elman   &            & \U{0.8442}   & 0.3586       &            & 0.9155       & 0.4381       \\*
		                                                                              &     & GRU     &            & 0.8433       & \U{0.3843}   &            & 0.9154       & 0.4400       \\*
		                                                                              &     & LSTM    &            & 0.8356       & 0.3514       &            & 0.9103       & \U{0.4411}   \\*
		\newembed
\embedding{\gls{surfperch}~\autocite{williams2024surfperch}}{1280}            & LR  &         & 0.8189     & \U{0.4055}   &              & 0.9087     & \U{0.4516}                  \\*
		                                                                              &     & MLP     &            & 0.7309       & 0.2545       &            & \U{0.9160}   & 0.4409       \\*
		                                                                              &     & Elman   &            & \U{0.8434}   & 0.3795       &            & 0.9117       & 0.4385       \\*
		                                                                              &     & GRU     &            & 0.8327       & 0.3646       &            & 0.9109       & 0.4441       \\*
		                                                                              &     & LSTM    &            & 0.8390       & 0.3669       &            & 0.9085       & 0.4432       \\*
		\newembed
\embedding{\gls{perchv2}~\autocite{Merrienboer2025perchv2}}{1536}             & LR  &         & 0.8020     & 0.3654       &              & 0.9319     & 0.4824                      \\*
		                                                                              &     & MLP     &            & 0.7976       & 0.2929       &            & \B\U{0.9355} & 0.4757       \\*
		                                                                              &     & Elman   &            & \B\U{0.8492} & \B\U{0.4220} &            & 0.9314       & 0.5077       \\*
		                                                                              &     & GRU     &            & 0.8453       & 0.3966       &            & 0.9312       & \B\U{0.5126} \\*
		                                                                              &     & LSTM    &            & 0.8400       & 0.3744       &            & 0.9276       & 0.5036       \\*
		\newembed
\embedding{\gls{aves-core}~\autocite{hagiwara2022aves}}{768}                  & LR  &         & 0.7995     & 0.2621       &              & 0.8794     & 0.2690                      \\*
		                                                                              &     & MLP     &            & \U{0.8210}   & \U{0.2789}   &            & \U{0.8907}   & 0.2767       \\*
		                                                                              &     & Elman   &            & 0.7297       & 0.1859       &            & 0.8648       & 0.3090       \\*
		                                                                              &     & GRU     &            & 0.7544       & 0.2407       &            & 0.8798       & \U{0.3543}   \\*
		                                                                              &     & LSTM    &            & 0.7559       & 0.2051       &            & 0.8635       & 0.3250       \\*
		\newembed
\embedding{\gls{aves-bio}~\autocite{hagiwara2022aves}}{768}                   & LR  &         & 0.8180     & 0.2799       &              & 0.8789     & 0.2836                      \\*
		                                                                              &     & MLP     &            & \U{0.8399}   & \U{0.3014}   &            & 0.8808       & 0.2935       \\*
		                                                                              &     & Elman   &            & 0.7691       & 0.2179       &            & 0.8685       & 0.3364       \\*
		                                                                              &     & GRU     &            & 0.8060       & 0.2586       &            & \U{0.8815}   & \U{0.3587}   \\*
		                                                                              &     & LSTM    &            & 0.7667       & 0.2295       &            & 0.8701       & 0.3232       \\*
		\newembed
\embedding{\gls{aves-nonbio}~\autocite{hagiwara2022aves}}{768}                & LR  &         & 0.8126     & 0.2574       &              & \U{0.8763} & 0.2752                      \\*
		                                                                              &     & MLP     &            & \U{0.8324}   & \U{0.2811}   &            & 0.8745       & 0.2804       \\*
		                                                                              &     & Elman   &            & 0.7700       & 0.2239       &            & 0.8536       & 0.3104       \\*
		                                                                              &     & GRU     &            & 0.8036       & 0.2397       &            & 0.8742       & \U{0.3400}   \\*
		                                                                              &     & LSTM    &            & 0.7850       & 0.2301       &            & 0.8568       & 0.3070       \\*
		\newembed
\embedding{\gls{aves-all}~\autocite{hagiwara2022aves}}{768}                   & LR  &         & 0.8083     & 0.2271       &              & \U{0.8726} & 0.2647                      \\*
		                                                                              &     & MLP     &            & \U{0.8301}   & \U{0.2475}   &            & 0.8722       & 0.2545       \\*
		                                                                              &     & Elman   &            & 0.7757       & 0.1927       &            & 0.8431       & 0.2884       \\*
		                                                                              &     & GRU     &            & 0.7867       & 0.2100       &            & 0.8696       & \U{0.3236}   \\*
		                                                                              &     & LSTM    &            & 0.7366       & 0.1830       &            & 0.8477       & 0.3024       \\*
		\newembed
\embedding{\gls{birdaves-biox-base}~\autocite{hagiwara2024birdaves}}{768}     & LR  &         & 0.8144     & 0.2284       &              & \U{0.8874} & 0.2828                      \\*
		                                                                              &     & MLP     &            & \U{0.8356}   & \U{0.2613}   &            & 0.8759       & 0.2772       \\*
		                                                                              &     & Elman   &            & 0.7592       & 0.2146       &            & 0.8567       & 0.3139       \\*
		                                                                              &     & GRU     &            & 0.7896       & 0.2336       &            & 0.8681       & \U{0.3411}   \\*
		                                                                              &     & LSTM    &            & 0.7690       & 0.2122       &            & 0.8658       & 0.3104       \\*
		\newembed
\embedding{\gls{birdaves-biox-large}~\autocite{hagiwara2024birdaves}}{1024}   & LR  &         & 0.8118     & 0.2399       &              & \U{0.8913} & 0.2853                      \\*
		                                                                              &     & MLP     &            & \U{0.8437}   & \U{0.2746}   &            & 0.8887       & 0.2852       \\*
		                                                                              &     & Elman   &            & 0.7719       & 0.2208       &            & 0.8517       & 0.2722       \\*
		                                                                              &     & GRU     &            & 0.7773       & 0.2291       &            & 0.8770       & \U{0.3274}   \\*
		                                                                              &     & LSTM    &            & 0.7639       & 0.2054       &            & 0.8542       & 0.3064       \\*
		\newembed
\embedding{\gls{birdaves-bioxn-large}~\autocite{hagiwara2024birdaves}}{1024}  & LR  &         & 0.8210     & 0.2525       &              & \U{0.8990} & 0.2859                      \\*
		                                                                              &     & MLP     &            & \U{0.8476}   & \U{0.2734}   &            & 0.8955       & 0.2939       \\*
		                                                                              &     & Elman   &            & 0.7539       & 0.1876       &            & 0.8541       & 0.3017       \\*
		                                                                              &     & GRU     &            & 0.7761       & 0.2145       &            & 0.8695       & \U{0.3194}   \\*
		                                                                              &     & LSTM    &            & 0.7623       & 0.2190       &            & 0.8592       & 0.3163       \\*
		\newembed
\embedding{\gls{animal2vec}~\autocite{Zimmermann2026Animal2vecMeerKAT}}{1024} & LR  &         & 0.7791     & 0.2075       &              & 0.8387     & 0.2048                      \\*
		                                                                              &     & MLP     &            & \U{0.8039}   & \U{0.2389}   &            & \U{0.8618}   & 0.2381       \\*
		                                                                              &     & Elman   &            & 0.7861       & 0.2218       &            & 0.8564       & 0.2559       \\*
		                                                                              &     & GRU     &            & 0.7695       & 0.2364       &            & 0.8435       & \U{0.2622}   \\*
		                                                                              &     & LSTM    &            & 0.7313       & 0.2115       &            & 0.8298       & 0.2435       \\*
\bottomrule
	\end{longtable}
\end{resultstable}
 \end{appendices}

\end{document}